\documentclass[11pt,a4paper]{article}
\usepackage[utf8]{inputenc}
\usepackage{amsmath}
\usepackage{amsthm}
\numberwithin{equation}{section}
\usepackage{amsfonts}
\usepackage{mathtools}
\usepackage{mathrsfs}
\usepackage{mathpazo}
\usepackage{multirow}
\usepackage{amssymb}
\usepackage{graphicx}
\usepackage{ifpdf}
\usepackage{braket}
\usepackage[dvipsnames]{xcolor}

\usepackage{cite}
\usepackage[bookmarks=true,colorlinks=true,linkcolor=black,citecolor=orange,urlcolor=orange,bookmarksnumbered]{hyperref}

\usepackage[left=2.50cm, right=2cm, top=2cm, bottom=3cm]{geometry}
\begin{document}
\thispagestyle{empty}

\begin{center}

\textit{\small in memory of Moritz Meisel}

\vspace{2.0truecm}

{\Large \bf On exceptional QP-manifolds}

\vspace{2.0truecm}

{David Osten}

\vspace{0.5truecm}

{\em Institute for Theoretical Physics (IFT), \\
University of Wroc\l aw, Poland \\
\vspace{0.2truecm}

Arnold Sommerfeld Center for Theoretical Physics, \\
Ludwig Maximilian University, Munich, Germany 
}

\vspace{0.5truecm}

{{\tt david.osten@uwr.edu.pl}}

\vspace{0.5truecm}
\end{center}

\begin{abstract}
The connection between two recent descriptions of tensor hierarchies -- namely, infinity-enhanced Leibniz algebroids, given by Bonezzi \& Hohm and Lavau \& Palmkvist, and the $p$-brane QP-manifolds constructed by Arvanitakis -- is made precise. This is done by presenting a duality-covariant version of latter.

The construction is based on the QP-manifold $T^\star[n]T[1]M \times \mathcal{H}[n]$, where $M$ corresponds to the internal manifold of a supergravity compactification and $\mathcal{H}[n]$ to a degree-shifted version of the infinity-enhanced Leibniz algebroid. Imposing that the canonical $Q$-structure on $T^\star[n] T[1] M$ is the derivative operator on $\mathcal{H}$ leads to a set of constraints. Solutions to these constraints correspond to $\frac{1}{2}$-BPS $p$-branes, suggesting that this is a new incarnation of a brane scan. Reduction w.r.t. to these constraints reproduces the known $p$-brane QP-manifolds. This is shown explicitly for the SL$(3) \times$SL$(2)$- and SL$(5)$-theories.

Furthermore, this setting is used to speculate about exceptional 'extended spaces' and QP-manifolds associated to Leibniz algebras. 
A proposal is made to realise differential graded manifolds associated to Leibniz algebras as non-Poisson subspaces (i.e. not Poisson reductions) of QP-manifolds similar to the above. 
Two examples for this proposal are discussed: generalised fluxes (including the dilaton flux) of O$(d,d)$ and the 3-bracket flux for the SL$(5)$-theory.
\end{abstract}

\pagebreak
\tableofcontents

\section{Introduction}
Gauge theories are a corner stone of modern physics. Typically, these are based on some underlying Lie algebra structure. The natural curiosity of a mathematical physicist but also certain evidence in string theory lead to the question, whether this can this be stretched to more general algebraic structures. The study of supergravity compactifications, in particular gauged supergravities \cite{deWit:2002vt,deWit:2004nw,deWit:2008ta,Bergshoeff:2009ph}, gave raise gauge theories based on Leibniz (or Loday) algebras \cite{Lavau:2017tvi,Kotov:2018vcz}, whereas Lie 3-algebras appeared in the study of 6d superconformal field theories \cite{Palmer:2013pka,Lavau:2014iva}. Former in turn require $p$-form hierarchies of field strengths, gauge fields and gauge transformations. 

The natural language for ungauged supergravity is the one of exceptional field theory and $E_{d(d)}$ generalised geometry \cite{Pacheco:2008ps,Berman:2010is,Berman:2011jh,Coimbra:2011ky,Berman:2012vc,Berman:2012uy,Coimbra:2012af,Hohm:2013pua,Lee:2014mla,Hohm:2014qga}, see \cite{Berman:2020tqn} for a recent review. Gauged supergravity, in addition, needs the introduction of tensor hierarchies and generalised Cartan calculus \cite{Hohm:2013vpa,Hohm:2013uia,Hohm:2014fxa,Abzalov:2015ega,Musaev:2015ces,Hohm:2015xna,Wang:2015hca}. This is well-understood from a target space perspective. A topic that did not get a lot of attention is how such Leibniz gauge fields and the associated tensor hierarchies couple or relate to world-volumes of $p$-branes. For conventional $p$-form gauge fields, it is clear that the natural object, to which they couple, are $p$-dimensional world-volumes, basically via the pull-back of theses gauge fields to the world-volume. This article aims to contribute in this direction.

$E_{d(d)}$-covariant $p$-brane world-volume theories have been constructed via actions \cite{Sakatani:2016sko,Blair:2017hhy,Sakatani:2017vbd,Arvanitakis:2018hfn,Blair:2019tww,Sakatani:2020umt} or via Hamiltonian formulations \cite{Duff:1990hn,Berman:2010is,Hatsuda:2012vm,Duff:2015jka,Sakatani:2020iad,Sakatani:2020wah,Osten:2021fil,Hatsuda:2022zpi}. Topological terms and topological field theories to such $p$-branes, that are the dynamical objects in type II and eleven-dimensional supergravities, have been considered, partly including tensor hierarchies \cite{Bouwknegt:2011vn,Arvanitakis:2018cyo,Chatzistavrakidis:2019seu,Arvanitakis:2021wkt,Arvanitakis:2022fvv}, but not in an $E_{d(d)}$-covariant way. In these cases the underlying $E_{d(d)}$-structure is rather hidden, and also the construction of the appearing manifolds is somewhat ad hoc. On the other hand, for O$(d,d)$ generalised geometry topological $\sigma$-models, namely Courant $\sigma$-models (topological strings), were established in \cite{Ikeda:2002wh,Roytenberg:2006qz,Cattaneo:2009zx}. These are based on the AKSZ-construction \cite{Alexandrov:1995kv} and concrete connections to generalised geometry were used for example in \cite{Chatzistavrakidis:2018ztm,Kokenyesi:2018ynq}. In related contexts for O$(d,d)$ generalised geometry in \cite{Deser:2014mxa,Heller:2016abk,Deser:2018oyg,Arvanitakis:2021lwo}, supergeometric methods were applied. One aim of this article is to present a way to derive similar objects for $p$-branes from an underlying object, an $E_{d(d)}$-covariant QP-manifold. 

For the purposes of this article a tensor hierarchy will be the sequence
\begin{equation}
	\mathcal{R}_1 \overset{\partial}{\longleftarrow} \mathcal{R}_2 \overset{\partial}{\longleftarrow} \mathcal{R}_3 \overset{\partial}{\longleftarrow} ...
\end{equation}
where $\mathcal{R}_p$ are representations of some underlying Lie algebra $\mathfrak{g}$, the Lie algebra of the duality groups $E_{d(d)}$ or O$(d,d)$. The representations relevant here have been collected in table \ref{table:Representations}.\linebreak $\partial:\ \mathcal{R}_p \rightarrow \mathcal{R}_{p-1}$ is a derivative operator with $\partial^2=0$. Also there is a product $\bullet: \ \mathcal{R}_p \times 
\mathcal{R}_q \rightarrow \mathcal{R}_{p+q}$. 

\begin{table}
\begin{align*}
\begin{array}{c|c|cccc}
\text{duality group} & \text{O}(d,d) & \text{SL}(3)\times \text{SL}(2)& \text{SL}(5) & \text{SO}(5,5) & E_{6(6)}  \\ \hline
\mathcal{R}_1 & \mathbf{2d} & \mathbf{(\bar{3},2)}& \mathbf{10} & \mathbf{16} & \mathbf{27}  \\
\mathcal{R}_2 & \mathbf{1} & \mathbf{(3,1)}& \mathbf{\bar{5}} & \mathbf{10} & \mathbf{\overline{27}}  \\
\mathcal{R}_3 & & \mathbf{(1,2)} & \mathbf{5} & \mathbf{\overline{16}} & \mathbf{1}  \\ 
\mathcal{R}_4 & & \mathbf{(\bar{3},1)}& \mathbf{\overline{10}} & \mathbf{1} &  \\
\mathcal{R}_5 &  &	\mathbf{(3,\bar{2})} & \mathbf{1} & &	\\
\mathcal{R}_6 & & \mathbf{1} & & &
\end{array}
\end{align*}
\caption{The representations relevant for the tensor hierarchy and generalised Cartan calculus of duality groups O$(d,d)$ and $E_{d(d)}$ for $d \leq 6$.}
\label{table:Representations}
\end{table}

Recently, Bonezzi \& Hohm \cite{Bonezzi:2019ygf,Bonezzi:2019bek} and Lavau \& Palmkvist \cite{Lavau:2019oja}, building on earlier results \cite{Palmkvist:2013vya,Cederwall:2018aab,Cederwall:2019bai,Cederwall:2019qnw}, understood the algebraic structure of a tensor hierarchy as a differential graded Lie algebra. There the tensor hierarchy is realised on a graded vector space
\begin{equation}
	\mathcal{H} = \mathcal{R}_1 \oplus \mathcal{R}_2 \oplus ... ,
\end{equation}
where the $\bullet$-product corresponds to the graded Lie bracket and the derivative $\partial$ is eponymous for the differential graded Lie algebras. In comparison, the construction by Arvanitakis \cite{Arvanitakis:2018cyo,Arvanitakis:2021wkt} is based on a QP-manifolds $\mathcal{M}$ of degree $n$, for which
\begin{equation}
\text{functions of degree}\  n-p \sim \mathcal{R}_p
\end{equation}
the graded Poisson structure ($P$-structure) corresponds to the $\bullet$-product and a Hamiltonian $Q$-structure to the derivative $\partial$. Apart from that, many features in these two constructions are strikingly similar. For example, a Leibniz product resp. a \textit{generalised Lie derivative} between objects in $\mathcal{R}_1$ can be defined as a derived bracket
\begin{equation}
	\mathcal{L}_{v_1} v_2 = \{ \{ \Theta, v_1 \} , v_2 \} .
\end{equation}
In section \ref{chap:Charges} it is presented how these two points of view -- the differential graded Lie algebra and QP-manifold ones -- can be connected. The central underlying object will be $\mathcal{M}^\star = T^\star[n] T[1] M \times \mathcal{H}[n]$. Relating the canonical Hamiltonian on $T^\star[n] T[1] M$ to be the derivative $\partial$ on $\mathcal{H}[n]$ leads to a \textit{hierarchy of of constraints} of the schematic form (suppressing the index structure):
\begin{equation}
\Phi^{(p)}: \ t^{(p)} \sim \xi \cdot t^{(p-1)}
\end{equation}
where $t^{(p)}$ are coordinates on $\mathcal{M}^\star$ corresponding to generators of $\mathcal{R}^{(p)}$ and $\xi$ are the fibre coordinates on $T[1]M$. Their explicit form is given in \eqref{eq:ChargeCondition>1}. Hence, the striven for QP-manifolds are
\begin{equation}
\mathcal{M} = T^\star[n] T[1] M \times \mathcal{H}[n] \big\vert_{\Phi^{(p)}}.
\end{equation}
Solutions to the constraints $\Phi^{(p)}$ will reproduce all the QP-manifolds constructed in \cite{Arvanitakis:2018cyo,Arvanitakis:2021wkt}. Consequently, this procedure could be considered as a new kind of \textit{brane scan}. This is shown explicitly for the low-dimensional exceptional field theories SL$(3) \times$SL$(2)$ and SL$(5)$. Some questions remain open already for $E_{6(6)}$ and even more fundamentally for $E_{7(7)}$ and $E_{8(8)}$, as a more intricate interplay of $T^\star[n] T[1] M$ and the tensor hierarchy or a more elaborate tensor hierarchy (then the one based on the representations in table \ref{table:Representations}) might need to be considered.

All of the above, in some sense, plays on the algebroid level, -- i.e. the derivatives $\partial$ of the tensor hierarchy are, in the examples here, related to derivatives w.r.t.\ to some underlying manifold. For Lie algebras, there is well-known statement that the $\mathfrak{g}[1]$ associated to a Lie algebra $\mathfrak{g}$ can be characterised by a cohomological vector field
\begin{equation}
	Q = {f^c}_{ab} \xi^a \xi^b \frac{\partial}{\partial \xi^c}
\end{equation}
because $Q^2 = 0$ is equivalent to the Jacobi identity of $\mathfrak{g}$. Section \ref{chap:Conjecture} mainly deals with the question whether such an equivalence can be found for Leibniz algebras as well. 

One aim is to generalise the following statement for the O$(d,d)$-setup \cite{Roytenberg:2002nu}. On $T^\star[2]T[1]M$, the underlying QP-manifold of the Courant $\sigma$-model, we can choose a different Hamiltonian
\begin{equation}
	\Theta = F_{ABC} \xi^A \xi^B \xi^C
\end{equation}
where $\xi^A = (\xi^a,z_a) = \eta^{AB} z_B$ are the degree 1 coordinates and the $F_{ABC}$ correspond to constant totally antisymmetric generalised fluxes of O$(d,d)$ generalised geometry. Indices are raised and lowered with the O$(d,d)$-metric $\eta$. With this Hamiltonian one has 
\begin{itemize}
	\item a derived bracket: $[z_A , z_B] = \{ \{ \Theta , z_A \} , z_B \} = {F^C}_{AB} z_C$
	\item the master equation $\{ \Theta , \Theta \} = 0 \quad \Leftrightarrow  \quad {F^E}_{[AB} F_{CD]E} = 0,$ which is the Bianchi identity for generalised fluxes \cite{Geissbuhler:2013uka}.
\end{itemize}
This is not very surprising, as the constant generalised fluxes from O$(d,d)$ generalised geometry correspond to structure constant of a Lie algebra. For $E_{d(d)}$ generalised fluxes this is not the case. There constant generalised fluxes correspond to structure constants of a Leibniz algebra.

It is easy to reproduce a derived bracket of the form $\{ \{ \Theta , z_A \} , z_B \} = {F^C}_{AB} z_C$ for a Leibniz algebra. But the second point -- a relation between the master equation and the Bianchi identity of generalised fluxes (the Leibniz identity) -- is more difficult. The conjecture put forward in section \ref{chap:Conjecture} is the following. The underlying manifold is $T^\star[n] \mathcal{H}[n]$. A natural choice of Hamiltonian which contains information of a Leibniz algebra is the following \eqref{eq:TwistedHamiltonianConjecture}
\begin{equation}
\Theta = {C^C}_{AB} \xi^A \bar{t}^{(1) B} t^{(1)}_C + S_{A,\mathcal{B}}{}^\mathcal{C} \xi^A \bar{t}^{(2)}_{\mathcal{C}} t^{(2) \mathcal{B}} .
\end{equation}
where $ {C^C}_{AB} = {F^C}_{[AB]}$ and $S_{A,\mathcal{B}}{}^\mathcal{C}$ correspond to the skew-symmetric and symmetric parts of the Leibniz algebra structure constants and $t^{(p)}_A,\bar{t}^{(p)A}$ are generators of $\mathcal{R}_p$ resp. $\bar{\mathcal{R}}_p$. $z_A$, $\xi^A$ are some functions on $\mathcal{R}_1$ resp. $\bar{\mathcal{R}}_1$. In order to proceed, two identifications need to be performed. Schematically, these are $t^{(1)} \sim z d$ and $t^{(1)} \sim \xi \cdot t^{(2)}$. Latter is similar to the constraints in section \ref{chap:Charges}. But here these conditions seem not to be implementable as constraints without introducing new negative degree coordinates. The results of all calculations, for example the master equation, are simply projected onto this constraint surface, i.e. without performing symplectic reduction
\begin{equation}
\{ \Theta , \Theta \} \big\vert_{\text{constraints}} = 0 \quad \Leftarrow \quad \text{Bianchi identity}. 
\end{equation}
The two examples on which this conjecture is demonstrated are the constant generalised fluxes for O$(d,d)$, including the dilaton flux $F_A$, and constant $\mathbf{Q}$-flux, corresponding to a Nambu bracket structure constants, in the SL$(5)$-theory.

On the way to this conjecture the question of exceptional \textit{extended space} in the context of the QP-manifolds pops up. The conclusion is that the inclusion of the extended coordinates seem to require the relaxation of either graded Jacobi identity of the $P$-structure or of the requirement of absence of no negative degree coordinates.

\pagebreak
\section{From enhanced Leibniz algebroids to $p$-brane QP-manifolds} \label{chap:Charges}

\subsection{The construction of the underlying QP-manifold and the constraint}
In \cite{Bonezzi:2019ygf,Bonezzi:2019bek} (infinity-)enhanced Leibniz algebras have been defined via a quadruple $(H,\bullet,\partial,\circ)$. For the purpose of this paper, we let $H$ be a graded vector space that is composed of a direct (graded) sum of representations $\mathcal{R}_i$ of a Lie algebra $\mathfrak{g}$, corresponding to a Lie group $G$
\begin{equation}
	H = \mathcal{R}_1 \oplus \mathcal{R}_2 \oplus \dots  \ .
\end{equation}
In this article, we aim for the application to the $T$- and low dimensional $U$-duality groups. 
$\mathcal{R}_1$ will be denoted by indices $K,L,M,...$, $\mathcal{R}_2$ by $\mathcal{K},\mathcal{L},\mathcal{M},...$ and, for a general $\mathcal{R}_p$, the indices are $\alpha_p , \beta_p , ...$. The other objects in the quadruple $(H,\bullet,\partial,\circ)$ are a product $\bullet: \ \mathcal{R}_p \times \mathcal{R}_q \rightarrow \mathcal{R}_{p+q}$, a Leibniz product $\circ: \ \mathcal{R}_1 \times \mathcal{R}_1$ and the derivative $\partial: \ \mathcal{R}_{p+1} \rightarrow \mathcal{R}_p$ with $\partial^2 = 0$. For the axiomatic definition, see \cite{Bonezzi:2019ygf,Lavau:2019oja,Bonezzi:2019bek}. Here, we take a different route and go one step further to realise such a structure as a QP-manifold, where
\begin{enumerate}
\item the generators $t_{\alpha_p}^{(p)}$ of the representations $\mathcal{R}_p$ are promoted to coordinates of a QP-manifold
\begin{equation}
\text{deg} \ t^{(p)} = n-p.
\end{equation}
So, in contrast to \cite{Lavau:2019oja,Bonezzi:2019bek}, the grading is chosen such that $\mathcal{R}_1$ has degree $n-1$.

\item the $\bullet$-product is identified with a degree $-n$ graded Poisson structure ($P$-structure) of functions of degree less $n$, 
\begin{equation}
\{ U^{(p)} , V^{(q)} \} = U^{(p)} \bullet V^{(q)}.
\end{equation}
So far, this degree $n$ is arbitrary. In the following, we call this graded Poisson manifold, given by $(H,\bullet)$, the ($n$-shifted) hierarchy algebra $\mathcal{H}[n]$. The $\mathcal{R}_1 \bullet \mathcal{R}_1 \rightarrow \mathcal{R}_2$ will often be specified by the $\eta$-symbol $(U \bullet V)_\mathcal{M} = \eta_{\mathcal{M},KL} U^K V^L$.

Let it be noted that $(\mathcal{H}[n],\bullet)$ is, in general, \textit{not} a graded symplectic manifold but, by construction, only a graded Poisson manifold.

\item the derivative $\partial: \ \mathcal{R}_{p+1} \rightarrow \mathcal{R}_p$ comes from a $Q$-structure, namely a Hamiltonian action $ \partial = \{\Theta , \ \}$. Where $\Theta$ is a degree $(n+1)$ function, satisfying $\{\Theta,\Theta\} = 0$.
 
\item the generalised Lie derivative for a $\Lambda = \Lambda^M t^{(1)}_M \in \mathcal{R}_1$ and $U = U^{\alpha_p} t^{(p)}_{\alpha_p} \in \mathcal{R}_p$ is given as the derived bracket
\begin{equation}
\mathcal{L}_\Lambda U = \{ \{ \Theta , \Lambda  \}, U \}. \label{eq:genLieDer}
\end{equation}
For $\Lambda_i \in \mathcal{R}_1$ this also gives the Leibniz product on the degree $(n-1)$-objects,
\begin{equation}
	\Lambda_1 \circ \Lambda_2 = \mathcal{L}_{\Lambda_1} \Lambda_2 = \{ \{ \Theta , \Lambda_1  \}, \Lambda_2 \}.
\end{equation}
The generalised Cartan calculus follows from the graded Jacobi identity: $\mathcal{L}_\Lambda U  = \{ \Lambda , \{ \Theta, U \} \} +  \{ \Theta , \{ \Lambda , U \} \} = \Lambda \bullet \partial U + \partial (\Lambda \bullet U)$.

\end{enumerate}
The properties of $(H,\bullet,\partial,\circ)$, as described in \cite{Bonezzi:2019ygf,Bonezzi:2019bek}, follow from the graded Jacobi identity of the underlying graded Poisson structure and the master equation $\{\Theta,\Theta \} = 0$. This is analogous to the treatment as differential graded Lie algebra in \cite{Lavau:2019oja,Bonezzi:2019bek}.
 
\paragraph{The underlying QP-manifold.} There are some obvious issues with the above:
\begin{itemize}
\item This differential graded manifold setting should be much more restrictive than the differential graded Lie algebra picture from \cite{Lavau:2019oja,Bonezzi:2019bek}. I.e. with such manifold we need to identify conventional products of functions, like $U^{(p)} V^{(q)}$, with some element $W^{(p+q-n)}$. This seems to imply a plethora of additional conditions, moreover even depending on $n$, none of with are a part of the original data of the infinity enhanced Leibniz algebroid.

\item So far, the Hamiltonian that should realise the chain of derivatives, $\partial = \{\Theta , \cdot \}$, is not defined. Also, it seems speculative that all of the derivatives $\partial_p: \ \mathcal{R}_{p+1} \rightarrow \mathcal{R}_p$ can come from the same Hamiltonian. For the differential graded Lie algebra this was explained in \cite{Lavau:2020pwa}.
\end{itemize}
Moreover, for the examples of the duality hierarchies considered here, the derivative $\partial$ is typically connected to some derivative on some underlying manifold, which did not appear so far:
\begin{equation}
\partial U^{(p)} \equiv {D_{\alpha_p}}^{\beta_{p-1},M} \partial_M U^{\alpha_p} t_{\beta_{p-1}}.
\end{equation}
For SL$(3) \times$SL$(2)$ and SL$(5)$ these ${D_{\alpha_p}}^{\beta_{p-1},M}$ have been collected in this article's conventions in appendix \ref{chap:conv}. One way to solve these issues will be based on the following QP-manifold
\begin{equation}
\mathcal{M}^\star = T^\star[n] T[1]M \times \mathcal{H}[n]
\end{equation}
over a smooth $D$-dimensional\footnote{For $E_{d(d)}$ tensor hierarchies $D$ should be $d$ resp. $(d-1)$ depending if we work on a solution of $M$-theory or IIb section. So, $M$ corresponds to the \textit{internal} manifold in the supergravity compactification context.} manifold. The relevant coordinates and their degrees are
\begin{equation}
\begin{array}{c|ccccccccc}
\text{degree} & ... & & 0 & 1 &  & ... & & n-1 & n \\ \hline \text{tensor hierarchy $\mathcal{H}[n]$} & ... & & t_{\alpha_n}^{(n)} & t_{\alpha_{n-1}}^{(n-1)} & & ... & & t_{\alpha_1}^{(1)} & \\
\text{canonical $T^\star[n]T[1]M$} & & & X^M & \xi^M & & & & z_M & \pi_M 
\end{array} \nonumber
\end{equation}
with the canonical brackets 
\begin{align*}
\{ X^M , \pi_N\} = \delta_N^M = \{z_N , \xi^M \}, \quad \{ t^{(p)} , t^{(q)} \} = t^{(p)} \bullet t^{(q)},
\end{align*}
latter given explicitly in appendix \ref{chap:conv} and the canonical Hamiltonian on $T^\star[n] T[1]M$,
\begin{equation}
\Theta = - \xi^M \pi_M .
\end{equation}
In order to keep $\mathfrak{g}$-covariance (and independence of the choice of section), the coordinates on $T^\star[n]T[1]M$, $(x^\mu , \xi^\mu, z_\mu , \pi_\mu)$, were extended to be full $\mathcal{R}_1$ resp. $\bar{\mathcal{R}}_1$ representations. On a $D$-dimensional solution to the section condition,
\begin{equation}
	\eta^{\mathcal{L}, MN} \partial_M \otimes \partial_N = 0, \label{eq:SectionCondition}
\end{equation}
i.e. $\partial_M = (\partial_\mu , 0)$, $\mu,\nu,... = 1,...D$, only the coordinates $(x^\mu , \xi^\mu, z_\mu , \pi_\mu)$ will appear explicitly in all the calculations that are relevant -- in particular, this is the case for the derived brackets $\{ \{ \Theta , u \} ,v \}$ for deg $u,v \leq n$, that really are of interest here. 

\paragraph{Charge conditions and the reduced QP-manifold.}
The interplay of the two ingredients\linebreak $T^\star[n] T[1]M$ and $\mathcal{H}[n]$ is twofold in order to address the above issues:
\begin{itemize}
 \item On a given solution of the section condition $\partial_M = (\partial_\mu ,0)$, the components $t_{\mu}^{(1)}$ are fixed to be $z_\mu$. In particular, $\{t^{1}_M , \xi^N \} \partial_N = \{z_M , \xi^N \} \partial_N = \partial_M$. As a constraint, we write:
\begin{equation}
	\Phi^{(1)}_\mu = t_{\mu}^{(1)} - z_\mu \approx 0
\end{equation}
or, in a duality covariant way,
\begin{equation}
	\eta^{\mathcal{L},MN} \Phi^{(1)}_M \partial_N = \eta^{\mathcal{L},MN} (t_M^{(1)} - z_M ) \partial_N \approx 0. \label{eq:ChargeCondition1}
\end{equation}

\item The key restriction on the interplay of the two ingredients $\mathcal{H}[n]$ and $T^\star[n]T[1]M$ is the derivative $\partial: \ \mathcal{R}_{p+1} \rightarrow \mathcal{R}_{p}$. I.e. $\{\Theta , U^{(p+1)} \} \equiv \partial U^{(p+1)} \in \mathcal{R}_p$. This leads to a hierarchy of constraints
\begin{equation}
\Phi^{(p) M}_{\alpha_p} \delta_M = \left( \xi^M t^{(p)}_{\alpha_p} - {D_{\alpha_p}}^{\beta_{p-1},M} t^{(p-1)}_{\beta_{p-1}} \right) \partial_M \approx  0 \label{eq:ChargeCondition>1}
\end{equation}
for $p > 1$, that fixes parts of the generators $t^{(p-1)}$ in terms of $t^{(p)}$ and $\xi$. Moreover, we expect them to fix $n$ as well. Following the literature \cite{Arvanitakis:2018hfn,Osten:2021fil}, where such conditions appeared first in the construction of duality covariant $p$-brane $\sigma$-models, these constraints are referred to as \textit{charge conditions}. For $G = E_{d(d)}$ (shown explicitly for $d\leq 4$), their solutions correspond to the $\frac{1}{2}$-BPS $p$-branes, as shown in the following subsection.
\end{itemize}
Hence, the QP-manifold $\mathcal{M} \subset \mathcal{M}^\star$ that we aimed for, fulfilling all the properties 1.-4. from the previous paragraph and consequently useful to define a infinity enhanced Leibniz algebroid, is 
\begin{equation}
\mathcal{M} = T^\star[n]T[1]M \times \mathcal{H}[n] \vert \big\vert_{\Phi^{(p)} \approx 0}
\end{equation}
i.e.\ the submanifold $\mathcal{M} \subset \mathcal{M}^\star$ being subject to a Poisson reduction w.r.t.\ the constraints $\Phi^{(p)} \approx 0$. 

\paragraph{Constraint algebra.} It should be checked whether the charge conditions lead to any secondary constraints or their application requires the Dirac procedure. For SL$(3) \times$SL$(2)$ and SL$(5)$, one finds that they are first class, i.e. they Poisson commute weakly. This is easy enough to show explicitly for any of the solutions of these constraints (the $p$-brane charges introduced in the next subsection). Without solving the constraints (and the applying the section first), it possible to be proven in a duality covariant manner for the SL$(5)$-theory. Using the standard conventions from appendix \ref{chap:conv}, we can write the constraints in the following explicit form,\footnote{only to be contracted with the derivatives in order to maintain SL$(5)$-covariance. For explicit solutions one has to fix a M-theory or type IIb section first, i.e. $\partial_{\mathcal{L} \mathcal{L}^\prime} = (\partial_\lambda , 0)$.}
\begin{align}
\eta^{\mathcal{L},MN} \Phi^{(1)}_M \partial_N &= 
\frac{1}{4} \epsilon^{\mathcal{L} \mathcal{M} \mathcal{M}^\prime \mathcal{N} \mathcal{N}^\prime} \left( t^{(1)}_{\mathcal{M} \mathcal{M}^\prime} - z_{\mathcal{M} \mathcal{M}^\prime} \right) \partial_{\mathcal{N} \mathcal{N}^\prime} \approx 0, \nonumber \\
\Phi^{(2) \mathcal{L} \mathcal{L}^\prime,\mathcal{M}} \partial_{\mathcal{L} \mathcal{L}^\prime} &= \left( \xi^{\mathcal{L} \mathcal{L}^\prime} t^{(2)\mathcal{M}} + \frac{1}{2} \epsilon^{\mathcal{L} \mathcal{L}^\prime \mathcal{M} \mathcal{N} \mathcal{N}^\prime} t^{(1)}_{\mathcal{N} \mathcal{N}^\prime} \right) \partial_{\mathcal{L} \mathcal{L}^\prime} \approx 0, \nonumber \\
{\Phi^{(3) \mathcal{L} \mathcal{L}^\prime}}_{\mathcal{M}} \partial_{\mathcal{L} \mathcal{L}^\prime} &= \left( \xi^{\mathcal{L} \mathcal{L}^\prime} t^{(3)}_{\mathcal{M}} + \delta^{\mathcal{L} \mathcal{L}^\prime}_{\mathcal{M} \mathcal{N}} t^{(2) \mathcal{N}} \right) \partial_{\mathcal{L} \mathcal{L}^\prime} \approx 0, \label{eq:ChargeConditionSL5}\\
 \Phi^{(4) \mathcal{L} \mathcal{L}^\prime,\mathcal{M} \mathcal{M}^\prime} \partial_{\mathcal{L} \mathcal{L}^\prime} &= \left( \xi^{\mathcal{L} \mathcal{L}^\prime} t^{(4)\mathcal{M}\mathcal{M}^\prime} + \epsilon^{\mathcal{N} \mathcal{M} \mathcal{M}^\prime \mathcal{L} \mathcal{L}^\prime} t^{(3)}_{\mathcal{N}} \right) \partial_{\mathcal{L} \mathcal{L}^\prime} \nonumber \approx 0, \\
 \Phi^{(5) \mathcal{L} \mathcal{L}^\prime} \partial_{\mathcal{L} \mathcal{L}^\prime} &= \left( \xi^{\mathcal{L} \mathcal{L}^\prime} t^{(5)} + t^{(4) \mathcal{L} \mathcal{L}^\prime} \right) \partial_{\mathcal{L} \mathcal{L}^\prime} \approx 0 . \nonumber
\end{align}
Using the canonical Poisson brackets on $M = T^\star[n]T[1]M \times \mathcal{H}[n]$ and the section condition, the corresponding constraint algebra is:
\begin{align}
\{\Phi^{(1)}_{\mathcal{M}\mathcal{M}^\prime} , \Phi^{(2) \mathcal{L} \mathcal{L}^\prime,\mathcal{N}} \} \partial_{\mathcal{L} \mathcal{L}^\prime} &= - \delta_{\mathcal{M} \mathcal{M}^\prime}^{\mathcal{N} \mathcal{N}^\prime} {\Phi^{(3) \mathcal{L}\mathcal{L}^\prime}}_{\mathcal{M}} \partial_{\mathcal{L} \mathcal{L}^\prime} \approx 0 \nonumber \\
\{\Phi^{(1)}_{\mathcal{M}\mathcal{M}^\prime} , {\Phi^{(3) \mathcal{L} \mathcal{L}^\prime}}_{\mathcal{N}} \} \partial_{\mathcal{L} \mathcal{L}^\prime} &= -\frac{1}{2} \epsilon_{\mathcal{K}\mathcal{K}^\prime \mathcal{M} \mathcal{M}^\prime \mathcal{N}} \Phi^{(4) \mathcal{L}\mathcal{L}^\prime,\mathcal{K}\mathcal{K}^\prime} \partial_{\mathcal{L} \mathcal{L}^\prime} \approx 0 \\
\{\Phi^{(2) \mathcal{K}\mathcal{K}^\prime,\mathcal{L}} , \Phi^{(2) \mathcal{M} \mathcal{M}^\prime,\mathcal{N}} \} \partial_{\mathcal{K} \mathcal{K}^\prime} \otimes \partial_{\mathcal{M} \mathcal{M}^\prime} &= - \left( \xi^{\mathcal{K} \mathcal{K}^\prime} \Phi^{(4) \mathcal{M} \mathcal{M}^\prime, \mathcal{LN}} + \epsilon^{\mathcal{K} \mathcal{K}^\prime \mathcal{L} \mathcal{N} \mathcal{R}} {\Phi^{(3) \mathcal{K} \mathcal{K}^\prime}}_{\mathcal{R}} \right) \partial_{\mathcal{K} \mathcal{K}^\prime} \otimes \partial_{\mathcal{M} \mathcal{M}^\prime} \nonumber \\
&\approx 0 . \nonumber
\end{align}
These constraints close into each other nicely. The other brackets, i.e. those involving $\Phi^{(4)}$ or $\Phi^{(5)}$ on either the left or right side, do not seem to commute weakly and generate secondary constraints, i.e. $t^{(5)} = 0$ and $t^{(4) M} \partial_M = 0$. As we will see in the next subsection, these conditions are fulfilled for any solution in the SL$(5)$-theory.\footnote{The physical reason for this is that there are no $4$-branes in SL$(5)$-theory, only lower dimensional $p$-branes. As we have that deg $t^{(5)} = p-4$, it would have negative degree for $p < 4$, hence $t^{(5)} = 0$. $t^{(4) M} \partial_M = 0$ is then implied by $\Phi^{(5)}$.} In other words, $t^{(5)} = 0$ and $t^{(4) M} \partial_M = 0$ are implied already  by the primary constraints \eqref{eq:ChargeCondition>1}.

\subsection{$E_{d(d)}$ and solutions for $d \leq 4$}
For SL$(5)$ in the standard conventions of appendix \ref{chap:conv}, the charge conditions \eqref{eq:ChargeConditionSL5} take the following form:
\begin{itemize}
\item M-theory section:
	\begin{align}
		 \xi^\nu t^{(2) \mu} &= t^{(1) \mu \nu}, \qquad \xi^\mu t^{(2)5}  = 0 \nonumber \\
		\xi^\mu t^{(3)}_5  &= t^{(2)\mu}, \qquad \xi^\mu t^{(3)}_\nu  =  t^{(2)}  \delta_\mu^\nu \\
		 \xi^{\mu} t^{(4) \nu} &= 0, \qquad \xi^\mu t^{(4)}_{\nu \nu^\prime} = \delta^\mu_{[\nu} t^{(3)}_{\nu^\prime ]} \nonumber
	\end{align}
	\item type IIb section:
		\begin{align}
			\xi_{\underline{\mu}} t^{(2)m} &= t^{(1)m}_{\underline{\mu}}, \qquad \xi_{\underline{\mu}} t^{(2)}_{\underline{\nu \nu}^\prime} = t^{(1)}_{\underline{\mu \nu \nu}^\prime} \nonumber \\
			\xi_{\underline{\mu}} t^{(3)}_m &= 0, \qquad \xi_{\underline{\mu}} t^{(3)}_{\underline{\nu}} = t^{(2)}_{\underline{\mu  \nu}} \quad ( \xi_{\underline{\mu}} t^{(3)\underline{\nu}\underline{\nu}^\prime} = \epsilon^{\underline{\lambda\nu}\underline{\nu}^\prime} t^{(2)}_{\underline{\mu  \lambda}}) \\
			\xi_{\underline{\mu}} t^{(4)}_{\underline{\nu}} &= 0, \qquad \xi_{\underline{\mu}} t^{(4) \underline{\nu}}_n = t^{(3)}_n \delta^{\underline{\nu}}_{\underline{\mu}}, \qquad \xi_{\underline{\mu}} t^{(4) \underline{\nu \nu}^\prime \underline{\nu}^{\prime \prime}} = \epsilon^{\underline{\nu \nu}^\prime \underline{\nu}^{\prime \prime}} t^{(3)}_{\underline{\mu}} \nonumber
		\end{align}
\end{itemize}
The corresponding solutions are:
\begin{eqnarray}
\begin{array}{c|c|c|c|c|c|c|c}
& & n & t^{(1)}_M & t^{(2) \mathcal{M}} & t^{(3)}_{\mathcal{M}} & t^{(4) M}  & t^{(5)} \\ \hline 
\text{M-theory} & \text{M2} & 3 & (z_\mu , q_{M2} \xi^\mu \xi^{\mu^\prime} ) & (q_{M2} \xi^\mu , 0) & (0 , q_{M2} ) & 0 & 0 \\ \hline
\text{type IIb} & \text{F1/D1} & 2 & (z^{\underline{\mu}}, q_m \xi_{\underline{\mu}},0) & (q_m , 0) & 0 & 0 & 0 \\
& \text{D3} & 4 & (z^{\underline{\mu}}, \tilde{q}_m \xi_{\underline{\mu}}, q_{D3} \xi_{\underline{\mu}} \xi_{\underline{\mu}^{\prime}} \xi_{\underline{\mu}^{\prime \prime}}) & (\tilde{q}_m , q_{D3} \xi_{\underline{\mu}} \xi_{\underline{\mu}^{\prime}}) & (0 , q_{D3} \xi_{\underline{\mu}}) & (0,0,q_{D3}) & 0
\end{array} \nonumber
\end{eqnarray}
For the D3-brane there is the remaining free coordinates $\tilde{q}_m$ with $\{ \tilde{q}_m ,\tilde{q}_n \} = \epsilon_{mn} q_{D3}$. Each solution (associated to a $p$-brane) is characterised by an element $q$ in $\mathcal{R}_{p+1}$, taking account to the fact that the charge conditions \eqref{eq:ChargeCondition>1} are linear in the generators $t^{(p)}$, i.e. solutions for same $n$ could be simply added. Moreover, there is always a solution with $t^{(p)} = 0$ for $p>1$ and $t^{(1)}_M = (z_\mu , 0)$. This solution corresponds to neglecting the hierarchy completely and will be dubbed \textit{point particle} solution in the following.

It is straightforward to check that the same solutions can be obtained from SL$(3)\times$SL$(2)$, starting from the charge conditions \eqref{eq:ChargeCondition>1}. Technical details, necessary for the calculation, can be found in appendix \ref{chap:conv}.

\paragraph{Some general comments. } The solutions correspond exactly to the QP-manifolds for the $\frac{1}{2}$-BPS branes, already found by Arvanitakis \cite{Arvanitakis:2018cyo,Arvanitakis:2021wkt}. Hence, this construction could be considered as new version of a \textit{brane scan} -- in particular one that, given a similar construction exists for $E_{d(d)}$ with $d\geq 7$, could give rise to world-volume incarnations of exotic branes, as well. The severely constrained form of the QP-manifold in comparison to the infinity-enhanced Leibniz algebra resp. differential graded Lie algebras might be surprising. One physical interpretation is the following: whereas latter were introduced at the level of the (gauged) supergravity, QP-manifolds are used describe (the topological part of) world-volume theories. These are not expected to contain the full information, as has been already observed on the level of actions \cite{Arvanitakis:2018hfn} and the current algebra \cite{Osten:2021fil}.  

So far, the charge conditions (in different settings) \eqref{eq:ChargeCondition>1} have appeared only as relations between $\mathcal{R}_1$- and $\mathcal{R}_2$-representations \cite{Arvanitakis:2018hfn,Osten:2021fil}. It was only possible to solve these equations, because an identification like $z_M \sim \eta_{\mathcal{K},LM} \xi^L \mathcal{Q}^{\mathcal{K}}$ was made. This requires an extended spaces, i.e. components of $\xi^L$ other than $\xi^\lambda$ appear.

The form of the reduced QP-manifold to a given $p$-brane is \textit{universal}, in sense that it does not depend w.r.t.\ which $E_{d(d)}$ it is defined (if dimensionality allows this $p$-brane). Of course, the dimension of $M$ must change. Typically, a solution of the charge condition \eqref{eq:ChargeCondition>1} does not fix the $n$ of the QP-structure. The above choice -- a $p$-brane -- is associated to the $n  = p +1$ is fixed by the characterising 'charges' $q_{M2},q_{D3},...$ to be degree $0$. In these solutions for $d \leq 4$, this ensures that
\begin{itemize}
\item \textit{functions of degree $(n-p)$ $\sim$ $\mathcal{R}_p$, for $p > 0$}, as observed in \cite{Arvanitakis:2018hfn}. This solves the above mentioned issue of the additional multiplicative structure (in comparison the a differential graded Lie algebra), i.e. that $U^{(p)} V^{(q)} \sim W^{(p+q-n)}$.

\item \textit{the hierarchy terminates}. There are no non-negative degrees. I.e. $t^{(p)} \equiv 0 $, for $p > n$ for all the found solutions.

\item the reduced manifold $\mathcal{M} \subset \mathcal{M}^\star$ actually becomes a graded symplectic manifold. In principle, the procedure does not impose this. In general, $\mathcal{M}$ is only a graded Poisson manifold. 
\end{itemize}
As a last remark on the applicability of the SL$(3) \times$SL$(2)$- and SL$(5)$-hierarchy to the present setup let us note that, in principle, there is a violation of the axioms of the graded Poisson structure of the hierarchies from the SL$(3) \times$SL$(2)$- and SL$(5)$-theories -- graded skew-symmetry resp. graded Jacobi identity for certain elements. These violations are not relevant on the $p$-brane charge solutions, see appendix \ref{chap:viol} for details, but are expected to appear at the end of the hierarchy as the dual graviton becomes relevant at $\mathcal{R}^{(9-d)}$ for $E_{d(d)}$.

\subsection{Issues with $E_{6(6)}$ and the 5-brane solutions.}The following convention for the $\bullet$-product and derivative $\partial$ for representations $\mathcal{R}_1 =\mathbf{27},\mathcal{R}_2 = \underline{\mathbf{27}},\mathcal{R}_3 = \mathbf{1}$ of $E_{6(6)}$ with generators $t^{(1)}_M , t^{(2)M} , t^{(3)}$ are used to define the $Q$- and $P$-structure:
\begin{align*}
	\{ t^{(1)}_M , t^{(1)}_N \} &= d_{LMN} t^{(2)L}, \quad \{ t^{(1)}_M , t^{(2)N} \} = \delta_M^N t^{(3)} \\
	\partial U^{(2)} &= d^{LMN} \partial_L U^{(2)}_M t^{(1)}_N, \qquad \partial U^{(3)} = \partial_M U^{(3)} t^{(2)M}
\end{align*}
where $d_{LMN}$ is the symmetric invariant of $E_{6(6)}$. Details on how this invariant looks in the index decompositions
\begin{align*}
\text{M-theory}: \quad V^M &= (v^\mu , v_{\mu_1 \mu_2} , v_{\mu_1 ... \mu_5} ) \\
\text{type IIb}: \quad V^M &= (v_{\underline{\mu}} , v^{\underline{\mu}}_m , v^{\underline{\mu}_1 \underline{\mu}_2 \underline{\mu}_3} , v^m_{\underline{\mu}_1 ... \underline{\mu}_5} )
\end{align*}
with $\mu,\nu,... = 1,...5$ resp. $\underline{\mu},\underline{\nu},... = 1,...,4$ and $m,n,...=1,2$, can be found, for example, in \cite{Sakatani:2017xcn,Osten:2021fil}. The corresponding solutions of the charge conditions \eqref{eq:ChargeCondition>1} (beyond those for the lower dimensional branes that also exist here and are the same as above) are:
\begin{small}
\begin{eqnarray}
\begin{array}{c|c|c|c|c|c|c}
& \mathcal{R}_1 & \mathcal{R}_2 & \mathcal{R}_3 \oplus \wedge^3 T^{\star}M & \wedge^2 T^{\star}M  & T^{\star}M & C_{\infty}(M)  \\ \hline 
 \text{M5}  & \left( \begin{array}{c} z_\mu \\ \tilde{q} \xi^{\mu_1} \xi^{\mu_2} \\ q_{M5} \xi^{\mu_1} ... \xi^{\mu_5} \end{array} \right) & \left( \begin{array}{c} \tilde{q} \xi^\mu \\ q_{M5} \xi^{\mu_1} ... \xi^{\mu_4} \\ 0 \end{array} \right) &  \left( \begin{array}{c} \tilde{q} \\  q_{M5} \xi^{\mu_1} \xi^{\mu_2} \xi^{\mu_3} \end{array} \right) & q_{M5} \xi^{\mu_1} \xi^{\mu_2} &  q_{M5} \xi^{\mu} & q_{M5}  \\ \hline
 \text{D5/NS5} & \left( \begin{array}{c} z_{\underline{\mu}} \\ \tilde{q}_m \xi_{\underline{\mu}} \\  0 \\ q^m  \xi_{\underline{\mu}_1} ... \xi_{\underline{\mu}_5} \end{array} \right) & \left( \begin{array}{c} 0 \\ \tilde{q}_m \\ 0 \\  q^m  \xi_{\underline{\mu}_1  } ... \xi_{\underline{\mu}_4} \end{array} \right) & \left( \begin{array}{c}
  0 \\ q^m \xi_{\underline{\mu}_1} \xi_{\underline{\mu}_2}  \xi_{\underline{\mu}_3}
\end{array} \right) & q^m \xi_{\underline{\mu}_1} \xi_{\underline{\mu}_2} & q^m \xi_{\underline{\mu}} & q^m
\end{array} \nonumber
\end{eqnarray}
\end{small} assuming $n=6$. $q^m$ and $q_{M5}$ are the NS5/D5 resp. M5 'charges', $\tilde{q}_m$ and $\tilde{q}$ are some degree $5$ resp. $4$ not fixed by solving the charge conditions \eqref{eq:ChargeCondition>1}. Latter have some similarities to the M2 and F1/D1 charges. Despite the fact that (for $q_{M5}$) this M5-solution takes the same form regarding the spaces of coordinates/functions as the QP-manifold associated to the M5 in \cite{Arvanitakis:2018cyo}, simply solving the charge condition does not lead to the same P-structure. E.g., here $\{\tilde{q},\tilde{q} \} = 0$, in comparison to $\{\tilde{q},\tilde{q} \} = 1$ in \cite{Arvanitakis:2018cyo}. In particular, this leads to inconsistencies like $\{ t^{(1) \mu \mu^\prime} , t^{(1) \nu \nu^\prime} \} = 0 \neq {d_M}^{\mu \mu^\prime \nu \nu^\prime} t^{(2)}$ for the M5-brane solution, and one arrives at two issues arising when applying this papers procedure for $E_{6(6)}$ that are left for future research:
\begin{itemize}
	\item For $E_{6(6)}$ the constraint algebra requires the Dirac procedure, as some of the constraints do not Poisson commute weakly, e.g. in the M-theory section
	\begin{equation}
		\{ {\Phi^{(2)\mu}_{\nu \nu^\prime}} , \Phi^{(3) \pi} \} \approx \delta^{\mu \pi}_{\nu \nu^\prime} \neq 0 .
	\end{equation}

	\item In order, to arrive at the M5-brane QP-manifold constructed by Arvanitakis \cite{Arvanitakis:2018cyo,Arvanitakis:2021wkt}, and also correctly reproduce, one needs to have $\{ t^{(3)} , t^{(3)} \} \neq 0$. Even, when applying the Dirac procedure, this is not possible because for the $E_{6(6)}$-hierarchy, as introduced above, $t^{(3)}$ is a central part of algebra.
\end{itemize}
A part of the solution to these problems might be \textit{not} to impose the constraint \linebreak $\Phi^{(3)} = \xi^M t^{(3)} \partial_M - t^{(2)M} \partial_M \approx 0$. Not imposing the constraint $\Phi^{(9-d)}$ for $E_{d(d)}$, so the one involving the end of the tensor hierarchy, would change nothing for the calculations for SL$(5)$ and SL$(3)\times$SL$(2)$, which would correspond to $d=3$ or $4$. Also, for 5-branes the appearing hierarchy of functions is longer than the length of the hierarchy, which ends here at $\mathcal{R}_3$. The other ones are part of the de Rham-complex $T[1]M$. In the end, a modified approach, similar to the $E_{8(8)}$-theory \cite{Hohm:2014fxa} where even the generalised Lie derivative needs to be adjusted in order to accomodate the dual graviton. Supposedly, one has to use a different hierarchy or there has to be an additional non-trivial interplay of the hierarchy and the de Rham-complex. 

\section{Towards differential graded manifolds from Leibniz algebras} \label{chap:Conjecture}

As commented on in \cite{Arvanitakis:2018cyo} one can consider more general degree $n+1$ functions as Hamiltonian for the resulting $p$-brane QP-manifolds. Namely, there are twists of the canonical Hamiltonian: $\Theta = \xi^\mu \pi_\mu + \textit{fluxes}$. This is generalising the statement for $T^\star[2] T[1] M$, that exact Courant algebroids are classified by $H^3(M)$, i.e. there the most general Hamiltonian looks like $\Theta = \xi^\mu \pi^\mu + H_{\mu \nu \rho} \xi^\mu \xi^\nu \xi^\rho$. In this article's language, the most general proper twist, considered in \cite{Arvanitakis:2018hfn}, takes the following form on the reduced QP-manifold $\mathcal{M} \subset \mathcal{M}^\star$ from the previous section:
\begin{equation}
\Theta = \xi^\mu \pi_\mu + {F^K}_{\mu \nu} \xi^\mu \xi^\nu t^{(1)}_K . \label{eq:TwistGeneral}
\end{equation}
The contribution from ${f^{\kappa}}_{\mu \nu} \xi^\mu \xi^\nu t^{(1)}_\kappa = {f^{\kappa}}_{\mu \nu} \xi^\mu \xi^\nu z_\kappa$ is not a proper twist and can be eliminated by coordinate redefinition: $\xi^\mu \mapsto \xi^{\prime \mu} = {e^\mu}_\nu(x) \xi^\nu$.

In this section first steps will be undertaken to find a duality-covariant version of \eqref{eq:TwistGeneral}, as the above expression obviously depends on the choice of section but also on the type of section solution, i.e.\ M-theory or type IIb. In particular, the inclusion of the 'non-geometric' fluxes and 'non-geometric' coordinates $\xi_{\mu \nu}$ is to be expected. Moreover, instead of going to the above twists of \textit{algebroid} structure, we will start here by considering Hamiltonians associated to \textit{algebra} structures.

\paragraph{Differential graded manifolds associated to Leibniz algebras.} Famously, the QP-manifold \linebreak $T^\star[2]\mathfrak{g}[1]$ associated to a Lie algebra $\mathfrak{g}$ has the following Hamiltonian $\Theta = {f^c}_{ab} \xi^a \xi^b z_c$ whose master equation $\{ \Theta , \Theta \} = 0$ is equivalent to the Jacobi identity of $\mathfrak{g}$ (with $\{ z_a , \xi^b \} = \delta_a^b$ and $\{\xi^a, \xi^b \} = 0 = \{z_a , z_b \}$. In analogy, a similar statement is true for the 'Courant algebra'
\begin{equation}
	\Theta = F_{ABC} \xi^A \xi^B \xi^C .
\end{equation}
with $\xi^{A} = (\xi^a , z_a)$ the degree 1 coordinates from $T^\star[2]T[1]M$ and constant $F_{ABC}$. With $\{ \xi^A , \xi^B \} = \eta^{AB}$ the master equation $\{ \Theta , \Theta \} = 0$ implies the Bianchi identity of the constant generalised fluxes $F_{ABC}$,
\begin{align}
{F^E}_{[AB} F_{CD]E} = 0.
\end{align}
In the following, the aim is to generalise such a construction to obtain a Hamiltonian $\Theta$ on some differential graded manifold which is characterised by a Leibniz algebra, s.t.
\begin{itemize}
	\item for some appropriate degree $(n-1)$ functions $z_A$ the Leibniz algebra is obtained as a derived bracket $\{\{\Theta , z_A \} ,z_B\} = {F^C}_{AB} z_C$
	\item the master equation $\{\Theta, \Theta \} = 0$ is implied by the Leibniz identity.
\end{itemize}
In comparison to the two example above the mixed symmetry of the  Leibniz algebra structure constants will pose a challenge.
  
\paragraph{$\mathbf{Q}$-flux in the SL$(5)$-theory.} The focus here will lie on Leibniz algebras corresponding to supergravity gaugings, and particularly those for the SL$(5)$ exceptional field theory. But the results will be valid generally. Particular attention will be given to the '$\mathbf{Q}$-flux', as the easiest non-standard twist. The generalised fluxes
\begin{equation}
\mathcal{L}_{E_A} E_B = {F^C}_{AB} E_C, \qquad {F^C}_{AB} = {E_N}^C \partial_A {E_B}^N - {E^C}_N \partial_B {E_C}^N + \eta^{\mathcal{E},CD} \eta_{\mathcal{E},BF} {E_N}^F \partial_D {E_A}^N \label{eq:generalised fluxes}
\end{equation}
associated to a frame ${E_B}^N \partial_N$ of the SL$(5)$-theory decompose into the $\mathbf{40}$ (trace-free ${\tau_{[\mathcal{A}\mathcal{A^\prime}\mathcal{B}]}}^{\mathcal{C}}$), $\mathbf{15}$ ($s_{( \mathcal{AA}^\prime )}$) and $\mathbf{10}$ ($a_{[\mathcal{A} \mathcal{A}^\prime ]}$) of SL$(5)$,
\begin{equation}
{F_{\mathcal{A}\mathcal{A}^\prime,\mathcal{B}\mathcal{B}^\prime}}^{\mathcal{C}\mathcal{C}^\prime} = 4 {F_{\mathcal{A}\mathcal{A}^\prime,[\mathcal{B}}}^{[\mathcal{C}} \delta^{\mathcal{C}^\prime]}_{\mathcal{B}^\prime ]}, \quad {F_{\mathcal{A}\mathcal{A}^\prime,\mathcal{B}}}^{\mathcal{C}} = {\tau_{\mathcal{A}\mathcal{A^\prime}\mathcal{B}}}^{\mathcal{C}} + \frac{1}{2} \delta^{\mathcal{C}}_{[\mathcal{A}} s_{\mathcal{A}^\prime] \mathcal{B}} - \frac{1}{6} \delta^{\mathcal{C}}_{\mathcal{B}} a_{\mathcal{A} \mathcal{A}^\prime } - \frac{1}{3} \delta^{\mathcal{C}}_{[\mathcal{A}} a_{\mathcal{A}^\prime] \mathcal{B}}.
\end{equation}
In the M-theory section these components decompose into the following \linebreak SL$(4)$-representations \cite{Blair:2014zba}:
\begin{equation}
	F_{abcd} \ (\mathbf{1}), \quad {f^c}_{ab} \ (\mathbf{20} \oplus \mathbf{4}), \quad {\mathbf{Q}_d}^{abc} \ (\mathbf{10} \oplus \mathbf{6}), \quad \mathbf{R}^{a,bcde} \ (\mathbf{4})
\end{equation}
and locally non-geometric contributions $\mathcal{T}^{a,b}$ and $\mathcal{U}_b$ that will play no further role here. Twists by the $F_4$- and the geometric $f$-flux are standard and can be included in the standard form \eqref{eq:TwistGeneral} and have been discussed already in \cite{Arvanitakis:2018cyo}.

What is dubbed $\mathbf{Q}$-flux here has recently been of considerable interest in the context of non-abelian U-duality, exceptional Drinfeld algebras and the construction of new uplifts to lower dimensional to 11-dimensional supergravity recently \cite{duBosque:2017dfc,Sakatani:2019zrs,Malek:2019xrf,Malek:2020hpo,Hlavaty:2020pfj,Blair:2020ndg,Sakatani:2020iad,Musaev:2020nrt,Gubarev:2020ydf,Musaev:2020bwm,Sakatani:2021eqo,Blair:2022gsx,Blair:2022ahh,Blair:2022ndb,Hassler:2022egz}. The trace-free resp. trace components of the $\mathbf{Q}$-flux correspond to the symmetric resp. skew-symmetric SL$(5)$-tensors:
\begin{equation}
{\mathbf{Q}_a}^{b_1 b_2 b_3} = {\tilde{f}_a}{}^{b_1 b_2 b_3} = \frac{1}{4} \epsilon^{b_1 b_2 b_3 b} (s_{ab} + 2 a_{ab})
\end{equation}
It has a natural interpretation as structure constants $\tilde{f}_a{}^{b_1 b_2 b_3}$ of a three-bracket. If the generalised fluxes \eqref{eq:generalised fluxes} are constant, the Bianchi identity they have to obey is the Leibniz identity:
\begin{equation}
{F^A}_{BE} {F^E}_{CD} = {F^A}_{CE} {F^E}_{BD} + {F^A}_{ED} {F^E}_{BC}. \label{eq:LeibnizId}
\end{equation}
If only this flux is turned on, the (Bianchi) Leibniz identity \eqref{eq:LeibnizId} implies that constant $\mathbf{Q}$-flux corresponds to structure constants of a three-bracket that fulfils the fundamental identity of a Nambu bracket \cite{Nambu:1973qe}: 
\begin{equation}
{\tilde{f}_a}{}^{d b_1 b_2} {\tilde{f}_d}{}^{c_1 c_2 c_3} = 3 {\tilde{f}_a}{}^{d [c_1 c_2} {\tilde{f}_d}{}^{c_3] b_1 b_2} \label{eq:FundamentalIdentity}
\end{equation}

\subsection{Twist by the Weitzenb\"ock connection}
The first try, how two realise generalised fluxes as a derived bracket, is simply by considering the derived bracket of generalised frames ${E_A}^M \in E_{d(d)}$, treated as elements of $\mathcal{R}_1$, in the setup of section \ref{chap:Charges}. Naturally, the generalised fluxes \eqref{eq:generalised fluxes} appear in generalised Lie derivative, the Leibniz bracket of the fields $t^{(1)}_A = {E_A}^M t^{(1)}_M$. In terms of some generalised frame fields in ${E_A}^M \in E_{d(d)}$. The Poisson brackets in $\mathcal{H}$ -- of $t^{(1)}_A$, $t^{(2)\mathcal{A}} = {E^\mathcal{A}}_\mathcal{M} t^{(2)\mathcal{M}}$, ... -- stay invariant\footnote{As usual, the generalised frame in different representations are related by the invariance condition of the $E_{d(d)}$-invariants. For example
\begin{equation}
{E_A}^M {E_B}^N {E_\mathcal{C}}^\mathcal{L} \eta_{\mathcal{L},MN} = \eta_{\mathcal{C},AB}. \nonumber
\end{equation}}, e.g. $\{ t^{(1)}_A , t^{(1)}_B \} = \eta_{\mathcal{C},AB} t^{(2) \mathcal{C}}$. The ones on $T^\star[n] T[1]M$ get naturally twisted by the Weitzenb\"ock connection ${\Omega_{A,B}}^C = (\partial_A {E_B}^M) {E_M}^C$:
\begin{equation}
\{ \pi_A , \pi_B \} = - 2 {\Omega_{[A,B]}}^C \pi_C, \quad \{ \pi_A , t^{(1)}_B \} = - {\Omega_{A,B}}^C t^{(1)}_C, \quad \{ \pi_A , \xi^B \} = {\Omega_{A,C}}^B \xi^C.
\end{equation}
The derived bracket will have the striven for form
\begin{align}
\{ \{ \Theta , t^{(1)}_A \} , t^{(1)}_B \} &= ( { \Omega_{A,B}}^C - {\Omega_{B,A}}^C ) t^{(1)}_C + {\Omega_{C,A}}^D \eta_{\mathcal{E},BD} \xi^C t^{(2)\mathcal{E}} \nonumber \\
&= (2 { \Omega_{[A,B]}}^C + \eta^{\mathcal{F},CE} \eta_{\mathcal{F},BD} {\Omega_{E,A}}^D ) t^{(1)}_C = {F^C}_{AB} t^{(1)}_C
\end{align}
where a charge condition relating $\mathcal{R}_1$ and $\mathcal{R}_2$ appeared in the form
\begin{equation}
{\Omega_{C,A}}^B  \xi^C t^{(2)\mathcal{E}} = {\Omega_{C,A}}^B  \eta^{\mathcal{E},CF} t^{(1)}_F .
\end{equation}
The Hamiltonian stays the same. So, in particular, the master equation $\{\Theta,\Theta\} = 0$ is fulfilled trivially. The consistency condition on the Weitzenb\"ock connection, and hence also on the generalised fluxes ${F^C}_{AB}$ is encoded in the Jacobi identity $\{\pi_{[A} , \{\pi_B , \pi_{C]} \} \} = 0$.

\paragraph{Going from algebroid to algebra.} One incentive was to have some QP-structure that represents a Leibniz algebra structure. In principle, this seems to be achieved here as the twist of the $P$-structure, characterised by an arbitrary generalised flux. But there are two points that indicate that this is not what we look for here. Coupling the tensor hierarchy to a whole algebroid $T^\star[n]T[1]M$ seemed fairly excessive and also the twist is characterised by the Weitzenb\"ock connection and not the generalised fluxes. Nevertheless, some algebraic identities can be derived like that. For example, one define the action of the derivative $\partial: \ \mathcal{R}_p \rightarrow R_{p-1}$.
\begin{equation}
\partial t^{(2)\mathcal{A}} = \{\Theta , t^{(2)\mathcal{A}} \} = \frac{1}{d-1} \eta^{\mathcal{A},BC} {S^D}_{BC} t^{(1)}_D \equiv S^{B,\mathcal{A}} t^{(1)}_B
\end{equation}
where ${S^D}_{BC} = {F^C}_{(AB)}$. This can be derived from $\{ \Theta , \{ t^{(1)}_A , t^{(1)}_B \} \} = 2 {S^C}_{AB} t^{(1)}_C$.

Naturally, the Weitzenb\"ock connection does not share the same properties as the generalised flux. Wheras we can describe a twist by the $F_4$ for example without problem, or geometric flux ${f^c}_{ab}$ to some Lie group. Nevertheless, the above construction shows that we can represent these known cases also by a twist of the Poisson structure and not only by additional terms in the Hamiltonian. 
For other standard components of ${F^C}_{AB}$, ${\Omega_{A,B}}^C$ is not necessarily constant even if ${F^C}_{AB}$ is, as shown in the following.

\paragraph{Example: the $\mathbf{Q}$-flux background.} Let us take the SL$(5)$-theory in the M-theory section $\partial_M = (\partial_m , 0)$, where we decomposed indices like $X^M = (x^m,x_{mm^\prime})$. Let us work in the non-geometric frame
\begin{equation}
{E_M}^A = \left( \begin{array}{cc} \delta_a^m & 0 \\ \beta^{aa^\prime m} & \delta^{aa^\prime}_{mm^\prime} \end{array} \right), \qquad \partial_C = (\partial_c , \beta^{cc^\prime d} \partial_d)
\end{equation}
and take $\beta^{abc} = {\tilde{f}_d}{}^{abc} x^d$. This has constant $\mathbf{Q}$-flux,
\begin{equation}
\mathbf{Q}_a{}^{b_1 b_2 b_3} = F^{b_3}{}_a{}^{b_1 b_2} = {\tilde{f}_a}{}^{b_1 b_2 b_3}
\end{equation}
and $\mathbf{R}$-flux, $\mathbf{R}^{b_1,a_1 a_2,b_2 b_3} = F^{b_1,a_1 a_2,b_2 b_3}$,
\begin{equation}
\mathbf{R}^{b_1,a_1 a_2,b_2 b_3}=  \left( {\tilde{f}_d}{}^{c a_1 a_2} {\tilde{f}_c}{}^{b_1 b_2 b_3} - 3 {\tilde{f}_d}{}^{c [b_1 b_2} {\tilde{f}_c}{}^{b_3] a_1 a_2} \right) x^d .
\end{equation}
Latter vanishes when the ${\tilde{f}_d}{}^{abc}$ are structure constants of a 3-bracket that satisfies the fundamental identity, $[{a_1} , {a_2} , [ {b_1} , {b_2} , {b_3} ]] = 3[[{a_1} , {a_2} , {b_1}] , {b_2} , {b_3} ] +$c.p. of $b_i$'s. The Weitzenb\"ock connection on the other hand is not constant. In particular it has the non-vanishing components
\begin{equation}
{\Omega_a}^{bb^\prime c} = \tilde{f}_a{}^{bb^\prime c}, \qquad {\Omega^{aa^\prime,bb^\prime c}} = \tilde{f}_e{}^{aa^\prime d}  \tilde{f}_d{}^{bb^\prime c} x^e.
\end{equation}
Hence, one needs to refer to coordinates of $M$ to describe this QP-structure and the aim to represent the Leibniz algebra without this can't be achieved by this naive and straightforward procedure.

\subsection{Speculation on extended exceptional QP-manifolds} 
So far, the underlying QP-manifold was phrased in a duality-covariant manner but the section condition \eqref{eq:SectionCondition} was imposed immediately. In principle, there are two ways how to generalise this. One is to really consider dependences on extended coordinates $X^M = (x^\mu , ...)$ -- this is not the aim here. Alternatively, one could consider extended degree $1$ coordinates $\xi^M$. This seems logical, as the constructions in \cite{Arvanitakis:2018cyo} and also in the present paper deal with completing the $\xi$'s conjugate coordinate $z_\mu$ into $\mathcal{R}_1$-representation $t^{(1)}_M$. Hence, we aim to introduce degree $1$ coordinates that are dual to some $t^{(1)}_M$. Such that for example in the M-theory decomposition $\{ t^{(1) \nu \nu^\prime} , \xi_{\mu \mu^\prime} \} = \delta^{\nu \nu^\prime}_{\mu \mu^\prime}$. In section \ref{chap:Conjecture} it is shown, that assuming $\{ t_M^{(1)} , \xi^N \} = \delta^N_M$ will be too restrictive for the purposes there.

Why should this be useful? One application of this QP-manifold picture is a derivation of the current algebra of world-volume theories \cite{Arvanitakis:2021wkt}. For these it was shown in \cite{Osten:2021fil} that the world-volume currents corresponding to $t^{(1)}_M$ could be written as exterior products of world-volume differentials of the coordinates, $\mathrm{d} X^M$, which correspond to $\xi^M$ in the QP-manifold language. 


\paragraph{What to generalise from O$(d,d)$.} For O$(d,d)$ extending $\xi^\mu$ to $\xi^M$ is natural. The degree $1$ coordinates $(\xi^\mu , z_\mu)$ on $T^\star[2] T[1] M$ can be phrased as $\xi^M = (\xi^\mu , z_\mu)$. With this one can consider Hamiltonians like $\Theta = F_{ABC} \xi^A \xi^B \xi^C$ corresponding to a \textit{constant generalised flux} background. 

One way to construct the corresponding Poisson structure from an underlying extended QP-manifold is the following. This procedure was performed in \cite{Blair:2014kla} on the level of the current algebra. Start with the canonical $T^*[2] T[1] M_{ext.}$ on the extended (doubled) base manifold $M_{ext.}$ with coordinates
\begin{equation}
\begin{array}{c|c|c|c|c}
& X^M & \xi^M & z_M & \pi_K \\ \hline
\text{degree} & 0 & 1 & 1 & 2
\end{array} \nonumber
\end{equation}
and canonical Poisson brackets $\{ \ , \ \}^\prime$:
\begin{equation}
\{ X^M , \pi_N\}^\prime = - \delta^M_N, \quad \{z_N , \xi^M \}^\prime = 2 \delta^M_N, \quad \{z_N , z_M \}^\prime = \{\xi^N , \xi^M \}^\prime = 0.
\end{equation}
The standard (unextended) QP-manifold is given by
\begin{itemize}
\item reducing to a solution of the section condition $\eta^{MN} \partial_M \otimes \partial_N = 0$

\item the (second class) constraint $z_M = \eta_{MN} \xi^N$
\end{itemize}
The corresponding Dirac brackets are
\begin{equation}
\{\xi^M , z_N \} = - \{z_N , \xi^M \} = \delta^M_N, \quad \{z_M , z_N \} = \eta_{NM}, \quad \{\xi^M , \xi^N \} = \eta^{MN} \label{eq:OddPstructure}
\end{equation}
Together with a solution of the section condition, so reducing $X^M \rightarrow (x^\mu, 0 )$/$\pi_M \rightarrow (\pi_\mu , 0)$. The result basically corresponds to the standard supermanifold $M$, $T^\star[2]T[1]M$.

\paragraph{Problems for the exceptional case.}
In analogy to the O$(d,d)$-case one would consider \linebreak $\mathcal{M}_{ext.}^\star = T^\star[n]T[1]M_{ext.} \times \mathcal{H}[n]$ with coordinates
\begin{equation}
\begin{array}{c|ccccccccc}
\text{degree} & ... & & 0 & 1 &  & ... & & n-1 & n \\ \hline  & ... & & t_{\alpha_n}^{(n)} & t_{\alpha_{n-1}}^{(n-1)} & & ... & & t_{\alpha_1}^{(1)} & \\
 & & & X^M & \xi^M & & & & z_M & \pi_M 
\end{array} \nonumber
\end{equation}
In analogy to \eqref{eq:ChargeCondition1} one identifies $t^{(1)}_M $ with $z_M$
\begin{equation}
\Phi^{(1)}_M = t_M^{(1)} - A_M^N z_N  \approx 0, \label{eq:ConstraintExt1}
\end{equation}
but now for the full $\mathcal{R}_1$-index, not only on a section. As will be shown later, $A$ is some linear operator preserving the SL$(4)$-structure. This constraint is second class: $\{ \Phi^{(1)}_M , \Phi_N^{(1)} \} = \eta_{\mathcal{L},MN} t^{(2) \mathcal{L}}$.\footnote{Setting $t^{(2)} = 0$ would be trivial, and correspond to the point particle, as discussed above.} Besides difficulties in inverting the operator it would lead to the necessity of negative degree coordinates. Latter can also be seen coming, in analogy to the O$(d,d)$ case, where $\{ \xi^M , \xi^N \} \neq 0$, the grading of the P-structure implies that deg$\{ \xi^M , \xi^N \}<0$ for $n>2$. The analogy to the constraint $\xi^M = \eta^{MN} z_N = \eta^{MN} t^{(1)}_N$ in the O$(d,d)$-case is the following ansatz that appeared already in \cite{Osten:2021fil}:
\begin{equation}
t^{(1)}_M = \frac{1}{p} \eta_{\mathcal{L},MN} \xi^N t^{(2) \mathcal{L}}, \label{eq:t1Ansatz}
\end{equation}
for some $p$ (which coincidently will correspond to the spatial dimension of the corresponding brane). In \cite{Osten:2021fil} it was also shown that this ansatz together with the constraint $\Phi^{(2)}$ from \eqref{eq:ChargeCondition>1}, in explicit terms as a constraint
\begin{equation}
\Phi^{(2) \mathcal{L}} = \left( \xi^N t^{(2) \mathcal{L}} - \frac{1}{p} \eta^{\mathcal{L},MN} \eta_{\mathcal{K},MP} \xi^P t^{(2) \mathcal{K}} \right) \partial_N  , \label{eq:ConstraintsExt>1}
\end{equation}
has the same kinds of solutions, as the hierarchy of constraints $\Phi^{(p)}$ presented in section \ref{chap:Charges}, i.e. the $\frac{1}{2}$-BPS $p$-branes. In general, the expression \eqref{eq:t1Ansatz} fixes more than the the constraints \eqref{eq:ChargeCondition>1}, namely also the components $t^{(1)}_\mu$. For example, a solution corresponding to the M2-brane could be given as follows:
\begin{equation}
	t^{(1)}_A = \left( -\frac{1}{2} \xi_{aa^\prime} \xi^{a^\prime} , - \xi^a \xi^{a^\prime} \right). \label{eq:M2ChargeExtended}
\end{equation}
Of course, this expression is fairly useless when aiming to compute Poisson brackets, as the illusive Poisson brackets $\{ \xi^M , \xi^N \}$ are needed. Two ways out of this problem and the problem of the second-class nature of the constraint \eqref{eq:ConstraintExt1} could be:
\begin{itemize}
	\item allowing for (new) negative degree coordinates.
	\item relaxing the requirements on the $P$-structure, namely one could allow for violations of the the Jacobi identity.
\end{itemize}
In the next subsection a proposal is made following latter option, as it is closer to the construction in the previous section.

\subsubsection{A proposal}
The proposal put forward here is that parts of the constraints  \eqref{eq:ConstraintExt1} \& \eqref{eq:ConstraintsExt>1} are \textit{not} being subjected to a Dirac procedure, but all calculations are done in an embedding (graded Poisson) space and only the results are only projected -- i.e. not Poisson reduced -- onto the constraint surface. Naturally, this resulting space would not be a graded Poisson manifold in general anymore and the Jacobi identity would be violated. Going the other way, the \textit{unconstrained} space would be a symplectic completion of the almost Poisson manifold that is conjectured here as the extended $p$-brane QP-manifold.

The proposal for a Leibniz algebra QP-manifold is based on an extended differential graded manifold 
\begin{equation}
\mathcal{M}^\star = T^\star [n] \mathcal{H}[n] = T^\star [n] (\mathcal{R}_1 [n-1] \times \mathcal{R}_2 [n-2] \times ...)
\end{equation}
with coordinates
\begin{equation}
\begin{array}{c|ccccccccc}
\text{degree} & ... & & 0 & 1 &  & ... & & n-1 & n \\ \hline  
& ... & & t_{\alpha_n}^{(n)} & t_{\alpha_{n-1}}^{(n-1)} & & ... & & t_{\alpha_1}^{(1)} & \\
& & & & \bar{t}_{\alpha_1}^{(1)} & & ... & & \bar{t}_{\alpha_{n-1}}^{(n-1)} & ... \\
 & & & & \xi^M & & & & z_M &
\end{array} \nonumber
\end{equation}
with only non-vanishing (canonical) Poisson brackets 
\begin{equation}
\{ \xi^M , z_N \} = \delta^M_N = \{\bar{t}^{(1)M} , t^{(1)}_N \}, \quad \{\bar{t}^{(2)}_\mathcal{M} , t^{(2) \mathcal{N}} \} = \delta^\mathcal{N}_\mathcal{M}, \quad \ldots \ .
\end{equation}
In contrast to $\mathcal{H}[n]$ from the previous subsection the $\bullet$-product structure is not imposed and conjectured to follow only on the constraint surface.\footnote{A generalisation to the algebroid case would be $T^\star[n]( T[1] M_{ext.} \times \mathcal{R}_1[n] \times \mathcal{R}_2[n] \times ...)$. } Hence, this proposal is \textit{not} a straight-forward generalisation of the setting in the previous section.

The dual pairs of coordinates, $\xi^M$, $z_M$ and $t^{(1)}_M$, $\bar{t}^{(1)M}$, are not independent of each other. For example in the M-theory section of the SL$(5)$-theory, the following ansatz for their relation preserving the SL$(4)$-structure of the M-theory section is made:
\begin{equation}
	t^{(1)}_M = (t_\mu^{(1)} , t^{(1) \mu \mu^\prime} )  = (A \ z_\mu , B \ z^{\mu \mu^\prime} ), \label{eq:IdentificationR1R1}
\end{equation}
$A,B$ being some real constants.\footnote{For consistency, one needs $\bar{t}^{(1)M } = (A^{-1} \xi^\mu , B^{-1} \xi_{\mu \mu^\prime} )$.} In analogy with $\Theta = \frac{1}{3 !} F_{ABC} \xi^A \xi^B \xi^C$, the Hamiltonian is proposed
\begin{equation}
\Theta \sim {C_{AB}}^C \xi^A \bar{t}^{(1) B} t^{(1)}_C + S_{A,\mathcal{B}}{}^\mathcal{C} \xi^A t^{(2)}_C t^{(2) \mathcal{B}} . \label{eq:TwistedHamiltonianConjecture}
\end{equation}
The symbols ${C_{AB}}^C$ and $S_{A,\mathcal{B}}{}^{\mathcal{C}}$ correspond to skew-symmetric resp. symmetric part of the ${F_{AB}}^C$: $C_{AB}{}^C = F^C{}_{[AB]}$, $S_{A,\mathcal{B}}{}^{\mathcal{C}} = \frac{1}{2} \eta_{\mathcal{B},AD} \eta^{\mathcal{C},EF} F^{D}{}_{EF}$.

The physical subspace $\mathcal{M} \subset \mathcal{M}^\star$ is conjectured to be the space subject to the constraints:
\begin{equation}
	t^{(1)}_A \approx \frac{1}{p} \eta_{\mathcal{C},AB} \xi^B t^{(2) \mathcal{C}}, \qquad \xi^c t^{(2) \mathcal{A}} \approx \eta^{\mathcal{A},Bc} t^{(1)}_B \label{eq:ChargeConditionExtended}
\end{equation}
and a suitable choice of identification \eqref{eq:IdentificationR1R1} of $(t^{(1)}_M$, $\bar{t}^{(1) M})$ with $(z_M , \xi^M)$. So, in contrast to the conservative construction in section \ref{chap:Charges}, here a \textit{charge solution} for a particular flux configuration will be characterised 
\begin{itemize}
	\item by a \textit{charge solution} from \eqref{eq:ChargeConditionExtended}. Solutions to these will be the same $p$-brane charges as in section \ref{chap:Charges}, with additional identifications of $t^{(1)}_\mu$ with products of the $\xi^M$ as in \eqref{eq:M2ChargeExtended}.
	
	\item additionally by an identification $t^{(1)}_M \sim {A_M}^N z_N$, as in \eqref{eq:IdentificationR1R1}, that will depend on the flux configuration at hand.
\end{itemize}
The conjecture is that, for each flux configuration ${F^C}_{AB}$, a identification like \eqref{eq:IdentificationR1R1} exists, s.t.
\begin{equation}
\{ \Theta , \Theta \} \vert_\mathcal{M} = 0 \quad \Leftarrow \quad \text{Leibniz identity for } {F^C}_{AB}.
\end{equation}
This conjecture is illustrated in two examples. The O$(d,d)$ generalised fluxes including the generalised flux $F_A$ associated to the representation $\mathcal{R}_2 = 1$, and the $\mathbf{Q}$-flux in SL$(5)$ exceptional field theory.

\subsubsection{The O$(d,d)$ generalised fluxes including the dilaton flux}
Surprisingly, this proposal gives also something new even in the O$(d,d)$-case. Applying the procedure can perform the reduction w.r.t. to the constraints \eqref{eq:ChargeConditionExtended} explicitly and the remaining free coordinates are:
\begin{equation}
\begin{array}{c|ccc}
\text{degree} & 0 & 1 & 2 \\ \hline
& \bar{h} & \xi^A  & h 
\end{array} \nonumber
\end{equation}
$h$ resp. $\bar{h}$ are the degree shifted generators of $\mathcal{R}_2 = \mathbf{1}$ resp. $\bar{\mathcal{R}}_2$. The non-vanishing Poisson brackets are $\{\xi^A , \xi^B \} = \eta^{AB}$, $\{\bar{h} , h \} = - \{ h , \bar{h} \} = 1$.

The Hamiltonian \eqref{eq:TwistedHamiltonianConjecture} takes the general form:
\begin{equation}
	\Theta = \frac{1}{3!} F_{ABC} \xi^A \xi^B \xi^C + F_A \xi^A \bar{h} h
\end{equation}
The Bianchi identities of constant O$(d,d)$ generalised fluxes including the dilaton flux $F_A$ \cite{Geissbuhler:2013uka},
\begin{equation}
	 {F^E}_{[AB} F_{CD]E} = 0, \quad F^A F_{ABC} = 0, \quad F_A F^A = 0,
\end{equation}
for the case that $F^{ABC} F_{ABC} = 0$,\footnote{An additional term $\frac{1}{6} F_{ABC} F^{ABC}$ which would be expected in the third Bianchi identity \cite{Geissbuhler:2013uka}, has to be neglected here. It is not clear at this stage why this can't be reproduced here. In \cite{Geissbuhler:2013uka}, it is stated that, from the supergravity perspective, this Bianchi identity comes from the inclusion of the RR-sector, which would explain the absence here.} can be reproduced from the master equation:
\begin{equation}
\{\Theta, \Theta \} = {F^E}_{AB} F_{CDE} \xi^A \xi^B \xi^C \xi^D + F^A F_{ABC} \xi^B \xi^C \bar{h} h + F_A F^A (\bar{h}h)^2 = 0.
\end{equation}

\subsubsection{Twist by the $\mathbf{Q}$-flux in SL$(5)$ M-theory section}
As a non-trivial example from the exceptional field theory hierarchies, let us consider the $\mathbf{Q}$-flux (three-bracket) in the SL$(5)$-theory in the M-theory. This is a popular example recently, as it appears as dual three-bracket structure constants in the discussion of exceptional Drinfeld algebras. 

The following underlying QP-manifold, based on the one of the M2-brane, is used.
\begin{equation}
\begin{array}{c|cccc}
\text{degree} & 0 & 1 & 2 & 3 \\ \hline
& & \xi^A , \bar{t}^{(1) A} & z_A , t^{(1)}_A & 
\end{array} \nonumber
\end{equation}
The $t^{(2)}$ or other generators are not need for the simple example considered here. The only non-vanishing Poisson brackets are $\{ z_A , \xi^B\} = \delta_A^B = \{ t^{(1)}_A , \bar{t}^{(1) B} \}$.

As mentioned earlier, the $\mathbf{Q}$-flux is characterised by the $s_{ab}$ and $a_{ab}$ alone, ${\mathbf{Q}_a}^{b_1 b_2 b_3} = {\tilde{f}_a}{}^{b_1 b_2 b_3}= \frac{1}{4} \epsilon^{b_1 b_2 b_3 b} (s_{ab} + 2 a_{ab})$. $s_{ab}$ corresponds to the traceless part of $\mathbf{Q}$, whereas ${\tilde{f}_a}{}^{a b c} = a^{bc}$. The simplest case is the presence of only $s_{ab} \neq 0$, then the candidate Hamiltonian \eqref{eq:TwistedHamiltonianConjecture} is:
\begin{align}
	\Theta_s &= {F_{AB}}^C \xi^A \bar{t}^{(1)B} t^{(1)}_C = - \frac{1}{2} {\tilde{f}_a}{}^{b_1 b_2 b_3} \left( \xi_{b_1 b_2} \xi_{b_3 c} z^{ac} + \left( 1 + \frac{A}{B} \right) \xi^a \xi_{b_1 b_2} z_{b_3} \right)
\end{align}
with $A,B$ being the constants from \eqref{eq:ChargeConditionExtended}. An identification $t^{(1)}_A = (A z_a , B z^{aa^\prime} )$ can be found namely $t^{(1)}_A = (z_a , \frac{1}{2} z^{aa^\prime} )$ (or any other with $A/B = 2$) such that the master equation becomes
\begin{equation}
	\{ \Theta , \Theta \} \sim \left( {\tilde{f}_a}{}^{d b_1 b_2} {\tilde{f}_d}{}^{c_1 c_2 c_3} - {\tilde{f}_a}{}^{d c_2 c_3} {\tilde{f}_d}{}^{c_1 b_1 b_2} \right) \xi_{b_1 b_2} \xi_{c_1 c_2} \xi_{c_3 e} \xi^a \xi^e = 0\label{eq:MasterEqQFlux}
\end{equation}
when projecting on the extended M2-brane charge solution $t^{(1)}_A = \left(-\frac{1}{2} \xi_{aa^\prime} \xi^{a^\prime} , - \xi^a \xi^{a^\prime} \right)$ \eqref{eq:M2ChargeExtended} of the 'would-be constraints' \eqref{eq:ChargeConditionExtended}. The master equation is fulfilled when ${\tilde{f}_a}{}^{b_1 b_2 b_3} $ are the structure constants of a Nambu three-bracket because \eqref{eq:MasterEqQFlux} is the fundamental identity \eqref{eq:FundamentalIdentity}.

As a last remark, let it be noted that the derived bracket picture breaks down. It will not be the case that $\{ \{ \Theta , t^{(1)}_A \} , t^{(1)}_B \} = {F^C}_{AB} t^{(1)}_C$ or $\{ \{ \Theta , z_A \} , z_B \} = {F^C}_{AB} z_C$ in general. If one expects that on the projected space the graded Jacobi identity does not hold anymore, several relations do not hold anymore -- for example between $Q^2 = 0$ and the master equation $\{ \Theta , \Theta \} = 0$:
\begin{equation}
\{ \Theta , \{ \Theta , X \} \} \sim \{ \{ \Theta , \Theta \} , X \} + \{ \Theta , \Theta , X \},
\end{equation}  
where $\{ \cdot , \cdot , \cdot \}$ is the Jacobiator of $\{ \cdot , \cdot \}$. In a similar way, the relation between the master equation and the consistency condition of a derived bracket breaks down. Alternatively, a different construction is thinkable in which the derived bracket property is still fulfilled, but $\{\Theta , \Theta \} \neq 0$. After all, the construction in this section is just a proposal.

\section{Outlook}

The construction of $E_{d(d)}$-covariant QP-manifolds that mediate between recent descriptions of (duality) tensor hierarchies and generalised Cartan calculus: namely the differential graded Lie algebra in \cite{Lavau:2019oja,Bonezzi:2019bek} -- for the description of the gauge structure from the supergravity perspective -- and QP-manifolds from \cite{Arvanitakis:2018cyo,Arvanitakis:2021wkt} -- for the derivation of topological terms and the current algebra of the $\frac{1}{2}$-BPS $p$-brane -- was successful. Some conceptual questions remain: the true power of this approach would be revealed if it could be applied to $E_{d(d)}$ for higher $d \geq 7$. Then one could expect QP-manifolds and a route to topological theories for exotic branes as well. But, already for $E_{6(6)}$ and the 5-branes, the hierarchy assumed here is too short and a more intricate interplay of the tensor hierarchy and the de-Rham complex is needed, as mentioned in section \ref{chap:Charges}. Moreover, for $E_{7(7)}$ and $E_{8(8)}$ there is a mixed-symmetry $\bullet$-product: $\mathcal{R}_1 \times \mathcal{R}_1 \rightarrow \mathcal{R}_2$. It is not clear how to realise this as a graded Poisson structure, similar to the construction in section \ref{chap:Charges}. The inclusion of compensator fields into the generalised Lie derivative or the tensor hierarchy in order to incorporate the dual graviton degress of freedom, analogously to \cite{Hohm:2018qhd}, in to this formalism might solve these issues.


Regarding the proposal made in section \ref{chap:Conjecture}, the central question is, is it possible to make a more natural construction including negative degree coordinates, s.t. the Dirac procedure can be performed and e.g. $\{ \Phi^{(1)}_K , \Phi^{(1)}_L \} = \eta_{\mathcal{M},KL} t^{(2) \mathcal{M}}$ could be inverted for the constraint $\Phi^{(1)}$ from \eqref{eq:ConstraintExt1}? Even when accepting this, the proposal suffers of a bit of arbitrariness. The necessity to introduce the freedom to choose the identification $t^{(1)} \sim z$ depending on the flux configuration one would like to reproduce, seems unnatural.
  
\subsection*{Acknowledgements}
The author thanks Athanasios Chatzistavrakidis and Saskia Demulder for discussions initiating this project, Chris Blair and Sylvain Lavau for comments on the first version and acknowledges Dieter L\"ust and the Faculty of Physics of Ludwig Maximilian University Munich for their extraordinary support in the year 2022.

This research is part of the project No. 2022/45/P/ST2/03995 co-funded by the National Science Centre and the European Union’s Horizon 2020 research and innovation programme under the Marie Sk\l odowska-Curie grant agreement no. 945339.
 
\includegraphics[width = 0.09 \textwidth]{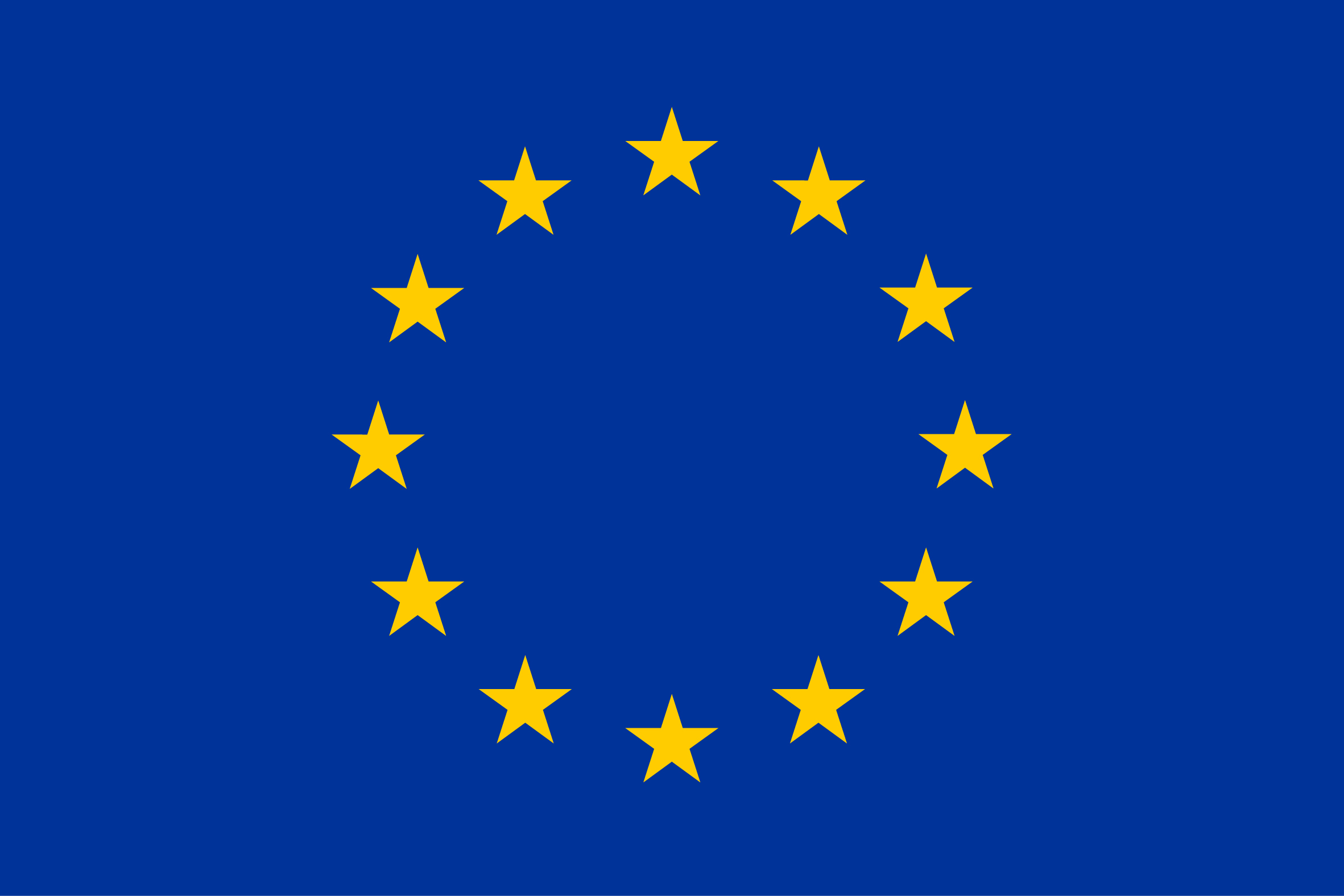} $\quad$
\includegraphics[width = 0.7 \textwidth]{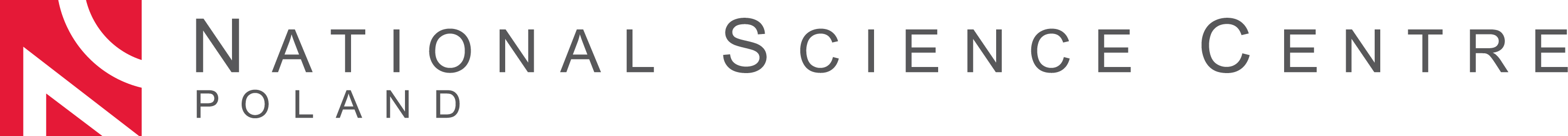}

  \appendix
  
  \section{Conventions for the SL$(3) \times$SL$(2)$- and SL$(5)$-tensor hierarchies} \label{chap:conv}
The conventions for the $\bullet$-product, $\bullet: \ \mathcal{R}_p \times \mathcal{R}_q \rightarrow \mathcal{R}_{p+q}$ and the derivation: $\partial: \ R_{p+1} \rightarrow R_p$ are given as there $\bullet$-product/graded Poisson bracket for generators of the $\mathcal{R}_p$ and the coefficients $D$ in $\partial U^{(p)} = {D_{\alpha_p}}^{\beta_{p-1},M} \partial_M U^{\alpha_p} t_{\beta_{p-1}} \in \mathcal{R}_{p-1}$. As in section 2, deg $\mathcal{R}_p = n-p$, where the graded Poisson bracket/the $\bullet$-product has degree $-n$. In general, conventions are used in which contractions of multi-indices $[\alpha_1 , ... , \alpha_p]$ receive factors of $\frac{1}{p!}$.

\subsection{SL$(3) \times$SL$(2)$}
$i,j,k,...$ denote fundamental SL$(3)$-indices, $\alpha,\beta,\gamma,...$ fundamental SL$(2)$-indices. For the generators $t^{(1)}_{i\alpha},t^{(2)i},t^{(3)}_\alpha,t^{(4)}_i,t^{(5)i\alpha},t^{(6)}$ of the representations $\mathcal{R}_1 , ... , \mathcal{R}_6$, after which the hierarchy cancels, the tensor hierarchy can be defined as follows:
\begin{align}
\{ t^{(1)}_{i\alpha} , t^{(1)}_{j \beta} \} &= \epsilon_{ijk} \epsilon_{\alpha \beta} t^{(2)k}, \quad \{ t^{(1)}_{i\alpha} , t^{(2)j} \} = \delta_i^j t^{(3)}_\alpha, \quad \{ t^{(1)}_{i\alpha} , t^{(3)}_\beta \} = \epsilon_{\alpha \beta} t^{(4)}_i, \nonumber \\
\{ t^{(1)}_{i\alpha} , t^{(4)}_j \} &= \epsilon_{ijk} \epsilon_{\alpha \beta} t^{(5)k\beta}, \quad \{ t^{(1)}_{i\alpha} , t^{(5)j \beta} \} = \delta_i^j \delta_\alpha^\beta t^{(6)}, \quad \{ t^{(2) i} , t^{(2)j} \} = \epsilon^{ijk} t^{(4)}_k,  \nonumber \\
\{ t^{(2) i} , t^{(3)}_\alpha \} &= \epsilon_{\alpha\beta} t^{(5)i\beta}, \quad \{ t^{(2) i} , t^{(4)}_j \} = \delta^i_j t^{(6)}, \quad \{ t^{(3)}_\alpha , t^{(3)}_\beta \} = \epsilon_{\alpha \beta} t^{(6)} \\
D^{k,i\alpha,j \beta} &= \epsilon^{ijk} \epsilon^{\alpha \beta}, \quad {D_{\alpha , i}}^{j \beta} = \delta_\alpha^\beta \delta_i^j, \quad {D_{i}}^{\alpha,j \beta} = \epsilon^{\alpha \beta} \delta_i^j,  \nonumber \\
 D^{i \alpha,j,k \gamma} &= \epsilon^{ijk} \epsilon^{\alpha \gamma},  \quad {D_{ \cdot , i\alpha}}^{j \beta} = \delta_i^j \delta_\alpha^\beta. \nonumber
\end{align}
All other Poisson brackets/$\bullet$-products vanish, or can be obtained by graded skew-symmetry. The section condition is $\epsilon^{ijk} \epsilon^{\alpha \beta} \partial_{i \alpha} \otimes \partial_{j \beta} = 0$. The M-theory section can be parametrised by $\partial_{i \alpha} = (\partial_{i1},\partial_{i2}) =  (\partial_{i1},0)$, the type IIb section by $\partial_{i \alpha} = (\partial_{1\alpha},\partial_{2\alpha},\partial_{3\alpha}) = (0,0,\partial_{3\alpha})$.

\subsection{SL$(5)$}
$\mathcal{K},\mathcal{L},...$ denote fundamental $\mathbf{5}$ indices, $K = [\mathcal{K}\mathcal{K}^\prime],L,M,...$ the $\mathbf{10}$-indices of SL$(5)$. Contractions of indices $[\mathcal{KK}^\prime],...$  get an extra factor $\frac{1}{2}$. For the generators $t^{(1)}_M,t^{(2)\,\mathcal{M}},t^{(3)}_\mathcal{M},t^{(4)M},t^{(5)}$ of the representations $\mathcal{R}_1 , ... , \mathcal{R}_5$, after which the hierarchy cancels, the tensor hierarchy can be defined as follows.
\begin{align}
\{ t^{(1)}_M , t^{(1)}_N \} &= \eta_{\mathcal{L},MN} t^{(2)\mathcal{L}}, \quad \{ t^{(1)}_{\mathcal{M}\mathcal{M}^\prime} , t^{(2) \mathcal{N}} \} = - \delta_{\mathcal{M} \mathcal{M}^\prime}^{\mathcal{N} \mathcal{L}} t^{(3)}_{\mathcal{L}} \nonumber \\
\{ t^{(1)}_{\mathcal{MM}^\prime} , t^{(3)}_{\mathcal{N}} \} &= \frac{1}{2} \epsilon_{\mathcal{N}\mathcal{L}\mathcal{L}^\prime \mathcal{M} \mathcal{M}^\prime} t^{(4) \mathcal{L} \mathcal{L}^\prime}, \quad  \{ t^{(1)}_M , t^{(4)N} \} = \delta_M^N t^{(5)} \nonumber \\
\{ t^{(2) \mathcal{M}} , t^{(2)\mathcal{N}}\} &= - t^{(4) \mathcal{MN}}, \quad \{ t^{(2)\mathcal{M}} , t^{(3)}_{\mathcal{N}} \} = \delta^\mathcal{M}_\mathcal{N} t^{(5)} \\
D^{\mathcal{K}, \mathcal{L} \mathcal{L}^\prime , \mathcal{M} \mathcal{M}^\prime} &= \epsilon^{\mathcal{KLL}^\prime \mathcal{MM}^\prime}, \quad {D_{\mathcal{L}\mathcal{M}}}^{\mathcal{NN}^\prime} = \delta^{\mathcal{NN}^\prime}_{\mathcal{LM}}, \nonumber \\
D^{\mathcal{K} \mathcal{K}^\prime , \mathcal{L},  \mathcal{M} \mathcal{M}^\prime} &= \epsilon^{\mathcal{KK}^\prime \mathcal{LMM}^\prime}, \quad {D_{\cdot , M}}^{N} = \delta_M^N . \nonumber
\end{align}
The section condition is $\epsilon^{\mathcal{KK}^\prime \mathcal{LL}^\prime M} \partial_{\mathcal{KK}^\prime} \otimes \partial_{\mathcal{LL}^\prime} = 0$. The M-theory section ($\kappa,\lambda,\mu,... = 1,...,4$) index decomposition is $V^{\mathcal{K}} = (V^\kappa,V^5)$, $V^K = V^{\mathcal{KK}^\prime} = (V^\kappa,V_{\kappa\kappa^\prime})$, with $\partial_K =  (\partial_{\kappa},0)$. In the type IIb section ($\underline{\kappa},\underline{\lambda},\underline{\mu},... = 1,2,3$, $k,l,m,... = 1,2$) the index decomposes as  $V^{\mathcal{K}} = (V^k,V_{\underline{\kappa \kappa}^\prime}) = (V^k,\epsilon_{\underline{\kappa \kappa}^\prime \underline{\lambda}} V^{\underline{\lambda}})$, $V^K = V^{\mathcal{KK}^\prime} = (V_{\underline{\kappa}},V^{\underline{\kappa}}_k,V^{\underline{\kappa \kappa}^\prime\underline{\kappa}^{\prime \prime}})$, with $\partial_{M} = (\partial^{\underline{\kappa}},0,0 = (\frac{1}{2} \epsilon^{\underline{\kappa \lambda \lambda}^\prime} \partial_{\underline{\lambda \lambda^\prime}}, 0 ,0)$.

\section{Remark on the violation of graded skew-symmetry and Jacobi identity} \label{chap:viol}
A graded Poisson structure $\{ \cdot , \cdot \}$ with degree $n$ is characterised, among a Leibniz rule, by 
\begin{align*}
&{} \text{graded skewsymmetry}: \quad \{ f,g \} = - (-1)^{(\text{deg}f -n)(\text{deg} g - n)} \{g,f\} \\
&{} \text{graded Jacobi identity}: \quad \{f,\{g,h\} \} = \{ \{f,g\} , h\} + (-1)^{(\text{deg}f -n)(\text{deg} g - n)} \{ g,\{f,h\} \}
\end{align*}
Let us note that, strictly speaking, the tensor hierarchies of the SL$(3) \times$SL$(2)$- and SL$(5)$-theories do violate these axioms:\begin{itemize}
\item For SL$(3) \times$SL$(2)$, there is a violation of the graded skew-symmetry for $C_i \in \mathcal{R}_3$:
\begin{equation}
C_1 \bullet C_2 \neq C_2 \bullet C_1, \quad \text{in general, but} \quad C_1 \bullet C_2 = - C_2 \bullet C_1 .
\end{equation}

\item For SL$(5)$, there is a violation of the graded Jacobi identity for $A \in \mathcal{R}_1$ and $B_i \in \mathcal{R}_2$
\begin{equation}
A \bullet (B_1 \bullet B_2 ) - (A \bullet B_1 ) \bullet B_2 + (A \bullet B_2 ) \bullet B_1 \neq 0 \qquad \text{in general}. 
\end{equation}

\end{itemize}
But, in both cases, these violations are \emph{not} relevant for the reduced $p$-brane QP-manifolds. After specifying to solutions to the charge conditions \eqref{eq:ChargeCondition>1}, discussed in section \ref{chap:Charges}, one notices that:
\begin{itemize}
\item for SL$(3) \times$SL$(2)$, on a solution of the constraints \eqref{eq:ChargeCondition>1}: $t^{(6)} = 0$, so 
\begin{equation}
\mathcal{R}^{(6)} \ni C_1 \bullet C_2 = 0
\end{equation}

\item for SL$(5)$, on a solution of the constraints \eqref{eq:ChargeCondition>1}: $t^{(5)} = 0$, so 
\begin{equation}
\mathcal{R}^{(5)} \ni A \bullet (B_1 \bullet B_2) =  A \bullet (B_1 \bullet B_2) = 0
\end{equation}
\end{itemize}
Hence, the resulting reduced spaces $\mathcal{M} \subset \mathcal{M}^\star$ are QP-manifolds.

\bibliographystyle{jhep}
\bibliography{References}

\providecommand{\href}[2]{#2}\begingroup\raggedright\begin{thebibliography}{10}

\bibitem{deWit:2002vt}
B.~de~Wit, H.~Samtleben, and M.~Trigiante, {\it {On Lagrangians and gaugings of
  maximal supergravities}},  {\em Nucl. Phys. B} {\bf 655} (2003) 93--126,
  [\href{http://arxiv.org/abs/hep-th/0212239}{{\tt hep-th/0212239}}].

\bibitem{deWit:2004nw}
B.~de~Wit, H.~Samtleben, and M.~Trigiante, {\it {The Maximal D=5
  supergravities}},  {\em Nucl. Phys. B} {\bf 716} (2005) 215--247,
  [\href{http://arxiv.org/abs/hep-th/0412173}{{\tt hep-th/0412173}}].

\bibitem{deWit:2008ta}
B.~de~Wit, H.~Nicolai, and H.~Samtleben, {\it {Gauged Supergravities, Tensor
  Hierarchies, and M-Theory}},  {\em JHEP} {\bf 02} (2008) 044,
  [\href{http://arxiv.org/abs/0801.1294}{{\tt arXiv:0801.1294}}].

\bibitem{Bergshoeff:2009ph}
E.~A. Bergshoeff, J.~Hartong, O.~Hohm, M.~Huebscher, and T.~Ortin, {\it {Gauge
  Theories, Duality Relations and the Tensor Hierarchy}},  {\em JHEP} {\bf 04}
  (2009) 123, [\href{http://arxiv.org/abs/0901.2054}{{\tt arXiv:0901.2054}}].

\bibitem{Lavau:2017tvi}
S.~Lavau, {\it {Tensor hierarchies and Leibniz algebras}},  {\em J. Geom.
  Phys.} {\bf 144} (2019) 147--189,
  [\href{http://arxiv.org/abs/1708.07068}{{\tt arXiv:1708.07068}}].

\bibitem{Kotov:2018vcz}
A.~Kotov and T.~Strobl, {\it {The Embedding Tensor, Leibniz\textendash{}Loday
  Algebras, and Their Higher Gauge Theories}},  {\em Commun. Math. Phys.} {\bf
  376} (2019), no.~1 235--258, [\href{http://arxiv.org/abs/1812.08611}{{\tt
  arXiv:1812.08611}}].

\bibitem{Palmer:2013pka}
S.~Palmer and C.~S\"amann, {\it {Six-Dimensional (1,0) Superconformal Models
  and Higher Gauge Theory}},  {\em J. Math. Phys.} {\bf 54} (2013) 113509,
  [\href{http://arxiv.org/abs/1308.2622}{{\tt arXiv:1308.2622}}].

\bibitem{Lavau:2014iva}
S.~Lavau, H.~Samtleben, and T.~Strobl, {\it {Hidden Q-structure and Lie
  3-algebra for non-abelian superconformal models in six dimensions}},  {\em J.
  Geom. Phys.} {\bf 86} (2014) 497--533,
  [\href{http://arxiv.org/abs/1403.7114}{{\tt arXiv:1403.7114}}].

\bibitem{Pacheco:2008ps}
P.~Pires~Pacheco and D.~Waldram, {\it {M-theory, exceptional generalised
  geometry and superpotentials}},  {\em JHEP} {\bf 09} (2008) 123,
  [\href{http://arxiv.org/abs/0804.1362}{{\tt arXiv:0804.1362}}].

\bibitem{Berman:2010is}
D.~S. Berman and M.~J. Perry, {\it {Generalized Geometry and M theory}},  {\em
  JHEP} {\bf 06} (2011) 074, [\href{http://arxiv.org/abs/1008.1763}{{\tt
  arXiv:1008.1763}}].

\bibitem{Berman:2011jh}
D.~S. Berman, H.~Godazgar, M.~J. Perry, and P.~West, {\it {Duality Invariant
  Actions and Generalised Geometry}},  {\em JHEP} {\bf 02} (2012) 108,
  [\href{http://arxiv.org/abs/1111.0459}{{\tt arXiv:1111.0459}}].

\bibitem{Coimbra:2011ky}
A.~Coimbra, C.~Strickland-Constable, and D.~Waldram, {\it {$E_{d(d)} \times
  \mathbb{R}^+$ generalised geometry, connections and M theory}},  {\em JHEP}
  {\bf 02} (2014) 054, [\href{http://arxiv.org/abs/1112.3989}{{\tt
  arXiv:1112.3989}}].

\bibitem{Berman:2012vc}
D.~S. Berman, M.~Cederwall, A.~Kleinschmidt, and D.~C. Thompson, {\it {The
  gauge structure of generalised diffeomorphisms}},  {\em JHEP} {\bf 01} (2013)
  064, [\href{http://arxiv.org/abs/1208.5884}{{\tt arXiv:1208.5884}}].

\bibitem{Berman:2012uy}
D.~S. Berman, E.~T. Musaev, D.~C. Thompson, and D.~C. Thompson, {\it {Duality
  Invariant M-theory: Gauged supergravities and Scherk-Schwarz reductions}},
  {\em JHEP} {\bf 10} (2012) 174, [\href{http://arxiv.org/abs/1208.0020}{{\tt
  arXiv:1208.0020}}].

\bibitem{Coimbra:2012af}
A.~Coimbra, C.~Strickland-Constable, and D.~Waldram, {\it {Supergravity as
  Generalised Geometry II: $E_{d(d)} \times \mathbb{R}^+$ and M theory}},  {\em
  JHEP} {\bf 03} (2014) 019, [\href{http://arxiv.org/abs/1212.1586}{{\tt
  arXiv:1212.1586}}].

\bibitem{Hohm:2013pua}
O.~Hohm and H.~Samtleben, {\it {Exceptional Form of D=11 Supergravity}},  {\em
  Phys. Rev. Lett.} {\bf 111} (2013) 231601,
  [\href{http://arxiv.org/abs/1308.1673}{{\tt arXiv:1308.1673}}].

\bibitem{Lee:2014mla}
K.~Lee, C.~Strickland-Constable, and D.~Waldram, {\it {Spheres, generalised
  parallelisability and consistent truncations}},  {\em Fortsch. Phys.} {\bf
  65} (2017), no.~10-11 1700048, [\href{http://arxiv.org/abs/1401.3360}{{\tt
  arXiv:1401.3360}}].

\bibitem{Hohm:2014qga}
O.~Hohm and H.~Samtleben, {\it {Consistent Kaluza-Klein Truncations via
  Exceptional Field Theory}},  {\em JHEP} {\bf 01} (2015) 131,
  [\href{http://arxiv.org/abs/1410.8145}{{\tt arXiv:1410.8145}}].

\bibitem{Berman:2020tqn}
D.~S. Berman and C.~D.~A. Blair, {\it {The Geometry, Branes and Applications of
  Exceptional Field Theory}},  {\em Int. J. Mod. Phys. A} {\bf 35} (2020),
  no.~30 2030014, [\href{http://arxiv.org/abs/2006.09777}{{\tt
  arXiv:2006.09777}}].

\bibitem{Hohm:2013vpa}
O.~Hohm and H.~Samtleben, {\it {Exceptional Field Theory I: $E_{6(6)}$
  covariant Form of M-Theory and Type IIB}},  {\em Phys. Rev.} {\bf D89}
  (2014), no.~6 066016, [\href{http://arxiv.org/abs/1312.0614}{{\tt
  arXiv:1312.0614}}].

\bibitem{Hohm:2013uia}
O.~Hohm and H.~Samtleben, {\it {Exceptional field theory. II. E$_{7(7)}$}},
  {\em Phys. Rev.} {\bf D89} (2014) 066017,
  [\href{http://arxiv.org/abs/1312.4542}{{\tt arXiv:1312.4542}}].

\bibitem{Hohm:2014fxa}
O.~Hohm and H.~Samtleben, {\it {Exceptional field theory. III. E$_{8(8)}$}},
  {\em Phys. Rev.} {\bf D90} (2014) 066002,
  [\href{http://arxiv.org/abs/1406.3348}{{\tt arXiv:1406.3348}}].

\bibitem{Abzalov:2015ega}
A.~Abzalov, I.~Bakhmatov, and E.~T. Musaev, {\it {Exceptional field theory:
  $SO(5,5)$}},  {\em JHEP} {\bf 06} (2015) 088,
  [\href{http://arxiv.org/abs/1504.01523}{{\tt arXiv:1504.01523}}].

\bibitem{Musaev:2015ces}
E.~T. Musaev, {\it {Exceptional field theory: $SL(5)$}},  {\em JHEP} {\bf 02}
  (2016) 012, [\href{http://arxiv.org/abs/1512.02163}{{\tt arXiv:1512.02163}}].

\bibitem{Hohm:2015xna}
O.~Hohm and Y.-N. Wang, {\it {Tensor hierarchy and generalized Cartan calculus
  in SL(3) \texttimes{} SL(2) exceptional field theory}},  {\em JHEP} {\bf 04}
  (2015) 050, [\href{http://arxiv.org/abs/1501.01600}{{\tt arXiv:1501.01600}}].

\bibitem{Wang:2015hca}
Y.-N. Wang, {\it {Generalized Cartan Calculus in general dimension}},  {\em
  JHEP} {\bf 07} (2015) 114, [\href{http://arxiv.org/abs/1504.04780}{{\tt
  arXiv:1504.04780}}].

\bibitem{Sakatani:2016sko}
Y.~Sakatani and S.~Uehara, {\it {Branes in Extended Spacetime: Brane
  Worldvolume Theory Based on Duality Symmetry}},  {\em Phys. Rev. Lett.} {\bf
  117} (2016), no.~19 191601, [\href{http://arxiv.org/abs/1607.04265}{{\tt
  arXiv:1607.04265}}].

\bibitem{Blair:2017hhy}
C.~D.~A. Blair and E.~T. Musaev, {\it {Five-brane actions in double field
  theory}},  {\em JHEP} {\bf 03} (2018) 111,
  [\href{http://arxiv.org/abs/1712.01739}{{\tt arXiv:1712.01739}}].

\bibitem{Sakatani:2017vbd}
Y.~Sakatani and S.~Uehara, {\it {Exceptional M-brane sigma models and
  $\eta$-symbols}},  {\em PTEP} {\bf 2018} (2018), no.~3 033B05,
  [\href{http://arxiv.org/abs/1712.10316}{{\tt arXiv:1712.10316}}].

\bibitem{Arvanitakis:2018hfn}
A.~S. Arvanitakis and C.~D. Blair, {\it {The Exceptional Sigma Model}},  {\em
  JHEP} {\bf 04} (2018) 064, [\href{http://arxiv.org/abs/1802.00442}{{\tt
  arXiv:1802.00442}}].

\bibitem{Blair:2019tww}
C.~D.~A. Blair, {\it {Open exceptional strings and D-branes}},  {\em JHEP} {\bf
  07} (2019) 083, [\href{http://arxiv.org/abs/1904.06714}{{\tt
  arXiv:1904.06714}}].

\bibitem{Sakatani:2020umt}
Y.~Sakatani and S.~Uehara, {\it {Born sigma model for branes in exceptional
  geometry}},  {\em PTEP} {\bf 2020} (2020), no.~7 073B05,
  [\href{http://arxiv.org/abs/2004.09486}{{\tt arXiv:2004.09486}}].

\bibitem{Duff:1990hn}
M.~J. Duff and J.~X. Lu, {\it {Duality Rotations in Membrane Theory}},  {\em
  Nucl. Phys.} {\bf B347} (1990) 394--419. [,210(1990)].

\bibitem{Hatsuda:2012vm}
M.~Hatsuda and K.~Kamimura, {\it {SL(5) duality from canonical M2-brane}},
  {\em JHEP} {\bf 11} (2012) 001, [\href{http://arxiv.org/abs/1208.1232}{{\tt
  arXiv:1208.1232}}].

\bibitem{Duff:2015jka}
M.~J. Duff, J.~X. Lu, R.~Percacci, C.~N. Pope, H.~Samtleben, and E.~Sezgin,
  {\it {Membrane Duality Revisited}},  {\em Nucl. Phys.} {\bf B901} (2015)
  1--21, [\href{http://arxiv.org/abs/1509.02915}{{\tt arXiv:1509.02915}}].

\bibitem{Sakatani:2020iad}
Y.~Sakatani and S.~Uehara, {\it {Non-Abelian $U$-duality for membranes}},  {\em
  PTEP} {\bf 2020} (2020), no.~7 073B01,
  [\href{http://arxiv.org/abs/2001.09983}{{\tt arXiv:2001.09983}}].

\bibitem{Sakatani:2020wah}
Y.~Sakatani, {\it {Extended Drinfel\textquoteright{}d algebras and non-Abelian
  duality}},  {\em PTEP} {\bf 2021} (2021), no.~6 063B02,
  [\href{http://arxiv.org/abs/2009.04454}{{\tt arXiv:2009.04454}}].

\bibitem{Osten:2021fil}
D.~Osten, {\it {Currents, charges and algebras in exceptional generalised
  geometry}},  {\em JHEP} {\bf 06} (2021) 070,
  [\href{http://arxiv.org/abs/2103.03267}{{\tt arXiv:2103.03267}}].

\bibitem{Hatsuda:2022zpi}
M.~Hatsuda, H.~Mori, S.~Sasaki, and M.~Yata, {\it {Gauged double field theory,
  current algebras and heterotic sigma models}},  {\em JHEP} {\bf 05} (2023)
  220, [\href{http://arxiv.org/abs/2212.06476}{{\tt arXiv:2212.06476}}].

\bibitem{Bouwknegt:2011vn}
P.~Bouwknegt and B.~Jurco, {\it {AKSZ construction of topological open p-brane
  action and Nambu brackets}},  {\em Rev. Math. Phys.} {\bf 25} (2013) 1330004,
  [\href{http://arxiv.org/abs/1110.0134}{{\tt arXiv:1110.0134}}].

\bibitem{Arvanitakis:2018cyo}
A.~S. Arvanitakis, {\it {Brane Wess-Zumino terms from AKSZ and exceptional
  generalised geometry as an $L_\infty$-algebroid}},  {\em Adv. Theor. Math.
  Phys.} {\bf 23} (2019), no.~5 1159--1213,
  [\href{http://arxiv.org/abs/1804.07303}{{\tt arXiv:1804.07303}}].

\bibitem{Chatzistavrakidis:2019seu}
A.~Chatzistavrakidis, L.~Jonke, D.~L\"ust, and R.~J. Szabo, {\it {Fluxes in
  Exceptional Field Theory and Threebrane Sigma-Models}},  {\em JHEP} {\bf 05}
  (2019) 055, [\href{http://arxiv.org/abs/1901.07775}{{\tt arXiv:1901.07775}}].

\bibitem{Arvanitakis:2021wkt}
A.~S. Arvanitakis, {\it {Brane current algebras and generalised geometry from
  QP manifolds. Or, \textquotedblleft{}when they go high, we go
  low\textquotedblright{}}},  {\em JHEP} {\bf 11} (2021) 114,
  [\href{http://arxiv.org/abs/2103.08608}{{\tt arXiv:2103.08608}}].

\bibitem{Arvanitakis:2022fvv}
A.~S. Arvanitakis, E.~Malek, and D.~Tennyson, {\it {Romans Massive QP
  Manifolds}},  {\em Universe} {\bf 8} (2022), no.~3 147,
  [\href{http://arxiv.org/abs/2201.07807}{{\tt arXiv:2201.07807}}].

\bibitem{Ikeda:2002wh}
N.~Ikeda, {\it {Chern-Simons gauge theory coupled with BF theory}},  {\em Int.
  J. Mod. Phys. A} {\bf 18} (2003) 2689--2702,
  [\href{http://arxiv.org/abs/hep-th/0203043}{{\tt hep-th/0203043}}].

\bibitem{Roytenberg:2006qz}
D.~Roytenberg, {\it {AKSZ-BV Formalism and Courant Algebroid-induced
  Topological Field Theories}},  {\em Lett. Math. Phys.} {\bf 79} (2007)
  143--159, [\href{http://arxiv.org/abs/hep-th/0608150}{{\tt hep-th/0608150}}].

\bibitem{Cattaneo:2009zx}
A.~S. Cattaneo, J.~Qiu, and M.~Zabzine, {\it {2D and 3D topological field
  theories for generalized complex geometry}},  {\em Adv. Theor. Math. Phys.}
  {\bf 14} (2010), no.~2 695--725, [\href{http://arxiv.org/abs/0911.0993}{{\tt
  arXiv:0911.0993}}].

\bibitem{Alexandrov:1995kv}
M.~Alexandrov, A.~Schwarz, O.~Zaboronsky, and M.~Kontsevich, {\it {The Geometry
  of the master equation and topological quantum field theory}},  {\em Int. J.
  Mod. Phys. A} {\bf 12} (1997) 1405--1429,
  [\href{http://arxiv.org/abs/hep-th/9502010}{{\tt hep-th/9502010}}].

\bibitem{Chatzistavrakidis:2018ztm}
A.~Chatzistavrakidis, L.~Jonke, F.~S. Khoo, and R.~J. Szabo, {\it {Double Field
  Theory and Membrane Sigma-Models}},  {\em JHEP} {\bf 07} (2018) 015,
  [\href{http://arxiv.org/abs/1802.07003}{{\tt arXiv:1802.07003}}].

\bibitem{Kokenyesi:2018ynq}
Z.~Kokenyesi, A.~Sinkovics, and R.~J. Szabo, {\it {AKSZ Constructions for
  Topological Membranes on $G_2$-Manifolds}},  {\em Fortsch. Phys.} {\bf 66}
  (2018), no.~3 1800018, [\href{http://arxiv.org/abs/1802.04581}{{\tt
  arXiv:1802.04581}}].

\bibitem{Deser:2014mxa}
A.~Deser and J.~Stasheff, {\it {Even symplectic supermanifolds and double field
  theory}},  {\em Commun. Math. Phys.} {\bf 339} (2015), no.~3 1003--1020,
  [\href{http://arxiv.org/abs/1406.3601}{{\tt arXiv:1406.3601}}].

\bibitem{Heller:2016abk}
M.~A. Heller, N.~Ikeda, and S.~Watamura, {\it {Unified picture of non-geometric
  fluxes and T-duality in double field theory via graded symplectic
  manifolds}},  {\em JHEP} {\bf 02} (2017) 078,
  [\href{http://arxiv.org/abs/1611.08346}{{\tt arXiv:1611.08346}}].

\bibitem{Deser:2018oyg}
A.~Deser and C.~S\"amann, {\it {Derived Brackets and Symmetries in Generalized
  Geometry and Double Field Theory}},  {\em PoS} {\bf CORFU2017} (2018) 141,
  [\href{http://arxiv.org/abs/1803.01659}{{\tt arXiv:1803.01659}}].

\bibitem{Arvanitakis:2021lwo}
A.~S. Arvanitakis, C.~D.~A. Blair, and D.~C. Thompson, {\it {A QP perspective
  on topology change in Poisson\textendash{}Lie T-duality}},  {\em J. Phys. A}
  {\bf 56} (2023), no.~25 255205, [\href{http://arxiv.org/abs/2110.08179}{{\tt
  arXiv:2110.08179}}].

\bibitem{Bonezzi:2019ygf}
R.~Bonezzi and O.~Hohm, {\it {Leibniz Gauge Theories and Infinity Structures}},
   {\em Commun. Math. Phys.} {\bf 377} (2020), no.~3 2027--2077,
  [\href{http://arxiv.org/abs/1904.11036}{{\tt arXiv:1904.11036}}].

\bibitem{Bonezzi:2019bek}
R.~Bonezzi and O.~Hohm, {\it {Duality Hierarchies and Differential Graded Lie
  Algebras}},  \href{http://arxiv.org/abs/1910.10399}{{\tt arXiv:1910.10399}}.

\bibitem{Lavau:2019oja}
S.~Lavau and J.~Palmkvist, {\it {Infinity-enhancing of Leibniz algebras}},
  {\em Lett. Math. Phys.} {\bf 110} (2020), no.~11 3121--3152,
  [\href{http://arxiv.org/abs/1907.05752}{{\tt arXiv:1907.05752}}].

\bibitem{Palmkvist:2013vya}
J.~Palmkvist, {\it {The tensor hierarchy algebra}},  {\em J. Math. Phys.} {\bf
  55} (2014) 011701, [\href{http://arxiv.org/abs/1305.0018}{{\tt
  arXiv:1305.0018}}].

\bibitem{Cederwall:2018aab}
M.~Cederwall and J.~Palmkvist, {\it {$L_{\infty }$ Algebras for Extended
  Geometry from Borcherds Superalgebras}},  {\em Commun. Math. Phys.} {\bf 369}
  (2019), no.~2 721--760, [\href{http://arxiv.org/abs/1804.04377}{{\tt
  arXiv:1804.04377}}].

\bibitem{Cederwall:2019bai}
M.~Cederwall and J.~Palmkvist, {\it {Tensor hierarchy algebras and extended
  geometry. Part II. Gauge structure and dynamics}},  {\em JHEP} {\bf 02}
  (2020) 145, [\href{http://arxiv.org/abs/1908.08696}{{\tt arXiv:1908.08696}}].

\bibitem{Cederwall:2019qnw}
M.~Cederwall and J.~Palmkvist, {\it {Tensor hierarchy algebras and extended
  geometry. Part I. Construction of the algebra}},  {\em JHEP} {\bf 02} (2020)
  144, [\href{http://arxiv.org/abs/1908.08695}{{\tt arXiv:1908.08695}}].

\bibitem{Roytenberg:2002nu}
D.~Roytenberg, {\it {On the structure of graded symplectic supermanifolds and
  Courant algebroids}},  in {\em {Workshop on Quantization, Deformations, and
  New Homological and Categorical Methods in Mathematical Physics}}, 3, 2002.
\newblock \href{http://arxiv.org/abs/math/0203110}{{\tt math/0203110}}.

\bibitem{Geissbuhler:2013uka}
D.~Geissb{\"{u}}hler, D.~Marqu{\'{e}}s, C.~Nu{\~{n}}ez, and V.~Penas, {\it
  {Exploring Double Field Theory}},  {\em JHEP} {\bf 06} (2013) 101,
  [\href{http://arxiv.org/abs/1304.1472}{{\tt arXiv:1304.1472}}].

\bibitem{Lavau:2020pwa}
S.~Lavau and J.~Stasheff, {\it {From Lie algebra crossed modules to tensor
  hierarchies}},  {\em J. Pure Appl. Algebra} {\bf 227} (2023) 107311,
  [\href{http://arxiv.org/abs/2003.07838}{{\tt arXiv:2003.07838}}].

\bibitem{Sakatani:2017xcn}
Y.~Sakatani and S.~Uehara, {\it {$\eta$-symbols in exceptional field theory}},
  {\em PTEP} {\bf 2017} (2017), no.~11 113B01,
  [\href{http://arxiv.org/abs/1708.06342}{{\tt arXiv:1708.06342}}].

\bibitem{Blair:2014zba}
C.~D.~A. Blair and E.~Malek, {\it {Geometry and fluxes of SL(5) exceptional
  field theory}},  {\em JHEP} {\bf 03} (2015) 144,
  [\href{http://arxiv.org/abs/1412.0635}{{\tt arXiv:1412.0635}}].

\bibitem{duBosque:2017dfc}
P.~du~Bosque, F.~Hassler, and D.~L\"ust, {\it {Generalized parallelizable
  spaces from exceptional field theory}},  {\em JHEP} {\bf 01} (2018) 117,
  [\href{http://arxiv.org/abs/1705.09304}{{\tt arXiv:1705.09304}}].

\bibitem{Sakatani:2019zrs}
Y.~Sakatani, {\it {$U$-duality extension of Drinfel\textquoteright{}d double}},
   {\em PTEP} {\bf 2020} (2020), no.~2 023B08,
  [\href{http://arxiv.org/abs/1911.06320}{{\tt arXiv:1911.06320}}].

\bibitem{Malek:2019xrf}
E.~Malek and D.~C. Thompson, {\it {Poisson-Lie U-duality in Exceptional Field
  Theory}},  {\em JHEP} {\bf 04} (2020) 058,
  [\href{http://arxiv.org/abs/1911.07833}{{\tt arXiv:1911.07833}}].

\bibitem{Malek:2020hpo}
E.~Malek, Y.~Sakatani, and D.~C. Thompson, {\it {E$_{6(6)}$ exceptional
  Drinfel\textquoteright{}d algebras}},  {\em JHEP} {\bf 01} (2021) 020,
  [\href{http://arxiv.org/abs/2007.08510}{{\tt arXiv:2007.08510}}].

\bibitem{Hlavaty:2020pfj}
L.~Hlavaty, {\it {Classification of 6D Leibniz algebras}},  {\em PTEP} {\bf
  2020} (2020), no.~7 071B01, [\href{http://arxiv.org/abs/2003.06164}{{\tt
  arXiv:2003.06164}}].

\bibitem{Blair:2020ndg}
C.~D.~A. Blair, D.~C. Thompson, and S.~Zhidkova, {\it {Exploring Exceptional
  Drinfeld Geometries}},  {\em JHEP} {\bf 09} (2020) 151,
  [\href{http://arxiv.org/abs/2006.12452}{{\tt arXiv:2006.12452}}].

\bibitem{Musaev:2020nrt}
E.~T. Musaev and Y.~Sakatani, {\it {Non-Abelian U duality at work}},  {\em
  Phys. Rev. D} {\bf 104} (2021), no.~4 046015,
  [\href{http://arxiv.org/abs/2012.13263}{{\tt arXiv:2012.13263}}].

\bibitem{Gubarev:2020ydf}
K.~Gubarev and E.~T. Musaev, {\it {Polyvector deformations in
  eleven-dimensional supergravity}},  {\em Phys. Rev. D} {\bf 103} (2021),
  no.~6 066021, [\href{http://arxiv.org/abs/2011.11424}{{\tt
  arXiv:2011.11424}}].

\bibitem{Musaev:2020bwm}
E.~T. Musaev, {\it {On non-abelian U-duality of 11D backgrounds}},  {\em
  Universe} {\bf 8} (2022) 276, [\href{http://arxiv.org/abs/2007.01213}{{\tt
  arXiv:2007.01213}}].

\bibitem{Sakatani:2021eqo}
Y.~Sakatani, {\it {Half-maximal extended Drinfel\textquoteright{}d algebras}},
  {\em PTEP} {\bf 2022} (2022), no.~1 013B14,
  [\href{http://arxiv.org/abs/2106.02041}{{\tt arXiv:2106.02041}}].

\bibitem{Blair:2022gsx}
C.~D.~A. Blair and S.~Zhidkova, {\it {Generalised U-dual solutions in
  supergravity}},  {\em JHEP} {\bf 05} (2022) 081,
  [\href{http://arxiv.org/abs/2203.01838}{{\tt arXiv:2203.01838}}].

\bibitem{Blair:2022ahh}
C.~D.~A. Blair, {\it {Non-isometric U-dualities}},  {\em JHEP} {\bf 09} (2022)
  115, [\href{http://arxiv.org/abs/2205.13019}{{\tt arXiv:2205.13019}}].

\bibitem{Blair:2022ndb}
C.~D.~A. Blair and S.~Zhidkova, {\it {Generalised U-dual solutions via ISO(7)
  gauged supergravity}},  {\em JHEP} {\bf 12} (2022) 093,
  [\href{http://arxiv.org/abs/2210.07867}{{\tt arXiv:2210.07867}}].

\bibitem{Hassler:2022egz}
F.~Hassler and Y.~Sakatani, {\it {All maximal gauged supergravities with
  uplift}},  \href{http://arxiv.org/abs/2212.14886}{{\tt arXiv:2212.14886}}.

\bibitem{Nambu:1973qe}
Y.~Nambu, {\it {Generalized Hamiltonian dynamics}},  {\em Phys. Rev.} {\bf D7}
  (1973) 2405--2412.

\bibitem{Blair:2014kla}
C.~D.~A. Blair, {\it {Non-commutativity and non-associativity of the doubled
  string in non-geometric backgrounds}},  {\em JHEP} {\bf 06} (2015) 091,
  [\href{http://arxiv.org/abs/1405.2283}{{\tt arXiv:1405.2283}}].

\bibitem{Hohm:2018qhd}
O.~Hohm and H.~Samtleben, {\it {The dual graviton in duality covariant
  theories}},  {\em Fortsch. Phys.} {\bf 67} (2019), no.~5 1900021,
  [\href{http://arxiv.org/abs/1807.07150}{{\tt arXiv:1807.07150}}].

\end{thebibliography}\endgroup
\end{document}